\newcount\driver
\newcount\mgnf \newcount\tipi
\newskip\ttglue
\def\TIPITOT{
\font\quattordicirm=cmr12 scaled\magstep2
\font\quattordicibf=cmbx12 scaled\magstep2
\font\quattordicisy=cmsy10 scaled\magstep2
\font\quattordiciex=cmex10 scaled\magstep2
\font\dodicirm=cmr12
\font\dodicii=cmmi12
\font\dodicisy=cmsy10 scaled\magstep1
\font\dodiciex=cmex10 scaled\magstep1
\font\dodiciit=cmti12
\font\dodicitt=cmtt12
\font\dodicibf=cmbx12 scaled\magstep1
\font\dodicisl=cmsl12
\font\ninerm=cmr9
\font\ninesy=cmsy9
\font\eightrm=cmr8
\font\eighti=cmmi8
\font\eightsy=cmsy8
\font\eightbf=cmbx8
\font\eighttt=cmtt8
\font\eightsl=cmsl8
\font\eightit=cmti8
\font\seirm=cmr6
\font\seibf=cmbx6
\font\seii=cmmi6
\font\seisy=cmsy6
\font\dodicitruecmr=cmr10 scaled\magstep1
\font\dodicitruecmsy=cmsy10 scaled\magstep1
\font\tentruecmr=cmr10
\font\tentruecmsy=cmsy10
\font\eighttruecmr=cmr8
\font\eighttruecmsy=cmsy8
\font\seventruecmr=cmr7
\font\seventruecmsy=cmsy7
\font\seitruecmr=cmr6
\font\seitruecmsy=cmsy6
\font\fivetruecmr=cmr5
\font\fivetruecmsy=cmsy5
\textfont\truecmr=\tentruecmr
\scriptfont\truecmr=\seventruecmr
\scriptscriptfont\truecmr=\fivetruecmr
\textfont\truecmsy=\tentruecmsy
\scriptfont\truecmsy=\seventruecmsy
\scriptscriptfont\truecmr=\fivetruecmr
\scriptscriptfont\truecmsy=\fivetruecmsy
\def \ottopunti{\def\rm{\fam0\eightrm}
\textfont0=\eightrm \scriptfont0=\seirm \scriptscriptfont0=\fiverm
\textfont1=\eighti \scriptfont1=\seii \scriptscriptfont1=\fivei
\textfont2=\eightsy \scriptfont2=\seisy \scriptscriptfont2=\fivesy
\textfont3=\tenex \scriptfont3=\tenex \scriptscriptfont3=\tenex
\textfont\itfam=\eightit \def\it{\fam\itfam\eightit}%
\textfont\slfam=\eightsl \def\sl{\fam\slfam\eightsl}
\textfont\ttfam=\eighttt \def\tt{\fam\ttfam\eighttt}%
\textfont\bffam=\eightbf \scriptfont\bffam=\seibf
\scriptscriptfont\bffam=\fivebf \def\bf{\fam\bffam\eightbf}%
\tt \ttglue=.5em plus.25em minus.15em
\setbox\strutbox=\hbox{\vrule height7pt depth2pt width0pt}%
\normalbaselineskip=9pt
\let\sc=\seirm \let\big=\eightbig \normalbaselines\rm
\textfont\truecmr=\eighttruecmr
\scriptfont\truecmr=\seitruecmr
\scriptscriptfont\truecmr=\fivetruecmr
\textfont\truecmsy=\eighttruecmsy
\scriptfont\truecmsy=\seitruecmsy
}\let\nota=\ottopunti}

\font\nineit=cmti10

\font\ninerm=cmr9

\font\ninerm=cmr9

\font\ninei=cmmi9

\font\ninesy=cmsy9

\font\ninebf=cmbx9

\font\ninett=cmtt9

\font\ninesl=cmsl9
\font\nineit=cmti9

\font\ninetruecmr=cmr9

\font\ninetruecmsy=cmsy9
\def \novepunti{\def\rm{\fam0\ninerm}%
\textfont0=\ninerm \scriptfont0=\seirm
\scriptscriptfont0=\fiverm %
\textfont1=\ninei \scriptfont1=\seii
\scriptscriptfont1=\fivei %
\textfont2=\ninesy \scriptfont2=\seisy
\scriptscriptfont2=\fivesy %
\textfont3=\tenex \scriptfont3=\tenex
\scriptscriptfont3=\tenex %
\textfont\itfam=\nineit
\def\it{\fam\itfam\nineit}
\textfont\slfam=\ninesl
\def\sl{\fam\slfam\ninesl}
\textfont\ttfam=\ninett
\def\tt{\fam\ttfam\ninett}
\textfont\bffam=\ninebf
\scriptfont\bffam=\seibf %
\scriptscriptfont\bffam=\fivebf
\def\bf{\fam\bffam\ninebf}
\tt \ttglue=.5em plus.25em
minus.15em %
\setbox\strutbox=\hbox{\vrule height7pt depth2pt
width0pt}

\normalbaselineskip=9pt
\let\sc=\seirm \let\big=\ninebig
\normalbaselines\rm %

\textfont\truecmr=\ninetruecmr

\scriptfont\truecmr=\seitruecmr

\scriptscriptfont\truecmr=\fivetruecmr

\textfont\truecmsy=\ninetruecmsy

\scriptfont\truecmsy=\seitruecmsy

}
\newfam\msbfam 
\newfam\truecmr 
\newfam\truecmsy 

\def\data{\number\day/\ifcase\month\or gennaio \or febbraio \or marzo
\or
aprile \or maggio \or giugno \or luglio \or agosto \or settembre
\or ottobre \or novembre \or dicembre \fi/\number\year;\,\the\time}
\def\dat{\number\day\,\ifcase\month\or {\rm gennaio} \or {\rm febbraio}
\or
{\rm marzo}
\or {\rm aprile} \or {\rm maggio} \or {\rm giugno} \or {\rm luglio} \or
{\rm agosto}
\or {\rm settembre} \or {\rm ottobre} \or {\rm novembre} \or {\rm
dicembre}
\fi\,\number\year}

\newcount\pgn \pgn=1
\def\foglio{\number\numsec:\number\pgn
\global\advance\pgn by 1}
\def\foglioa{A\number\numsec:\number\pgn
\global\advance\pgn by 1}

\global\newcount\numsec\global\newcount\numfor
\gdef\profonditastruttura{\dp\strutbox}

\def\senondefinito#1{\expandafter\ifx\csname#1\endcsname\relax}

\def\SIA #1,#2,#3 {\senondefinito{#1#2}%
\expandafter\xdef\csname #1#2\endcsname{#3}\else
\write16{???? ma #1,#2 e' gia' stato definito !!!!} \fi}

\def\etichetta(#1){(\veroparagrafo.\veraformula)%
\SIA e,#1,(\veroparagrafo.\veraformula) %
\global\advance\numfor by 1%
\write15{\string\FU (#1){\equ(#1)}}%
}

\def\FU(#1)#2{\SIA fu,#1,#2 }

\def\etichettaa(#1){(A.\veraformula)%
\SIA e,#1,(A.\veraformula) %
\global\advance\numfor by 1%
\write15{\string\FU (#1){\equ(#1)}}%
}

\def\getichetta(#1){Fig. \verafigura
\SIA e,#1,{\verafigura} %
\global\advance\numfig by 1%
\write15{\string\FU (#1){\equ(#1)}}%
\write16{ Fig. \equ(#1) ha simbolo #1 }}

\newdimen\gwidth

\def\BOZZA{
\def\alato(##1){%
{\vtop to \profonditastruttura{\baselineskip
\profonditastruttura\vss
\rlap{\kern-\hsize\kern-1.2truecm{$\scriptstyle##1$}}}}}
\def\galato(##1){\gwidth=\hsize \divide\gwidth by 2%
{\vtop to \profonditastruttura{\baselineskip
\profonditastruttura\vss
\rlap{\kern-\gwidth\kern-1.2truecm{$\scriptstyle##1$}}}}}
}

\def\alato(#1){}
\def\galato(#1){}

\def\veroparagrafo{\number\numsec}\def\veraformula{\number\numfor}
\def\verafigura{\number\numfig}

\def\geq(#1){\getichetta(#1)\galato(#1)}
\def\Eq(#1){\eqno{\etichetta(#1)\alato(#1)}}
\def\eq(#1){\etichetta(#1)\alato(#1)}
\def\Eqa(#1){\eqno{\etichettaa(#1)\alato(#1)}}
\def\eqa(#1){\etichettaa(#1)\alato(#1)}
\def\eqv(#1){\senondefinito{fu#1} #1
\write16{#1 non e' (ancora) definito}%
\else\csname fu#1\endcsname\fi}
\def\equ(#1){\senondefinito{e#1}\eqv(#1)\else\csname e#1\endcsname\fi}

\def\etichettab(#1){(B.\veraformula)%
\SIA e,#1,(B.\veraformula) %
\global\advance\numfor by 1%
\write15{\string\FU (#1){\equ(#1)}}%
}
\def\etichettac(#1){(C.\veraformula)%
\SIA e,#1,(C.\veraformula) %
\global\advance\numfor by 1%
\write15{\string\FU (#1){\equ(#1)}}%
}

\def\Eqb(#1){\eqno{\etichettab(#1)\alato(#1)}}
\def\Eqc(#1){\eqno{\etichettac(#1)\alato(#1)}}
\def\include#1{
\openin13=#1.aux \ifeof13 \relax \else
\input #1.aux \closein13 \fi}

\openin14=\jobname.aux \ifeof14 \relax \else
\input \jobname.aux \closein14 \fi
\openout15=\jobname.aux



 \def\\{\noindent}

\let \pt=\partial

\let \n\nabla

\def\tende#1{\vtop{\ialign{##\crcr\rightarrowfill\crcr
\noalign{\kern-1pt\nointerlineskip}
\hskip3.pt${\scriptstyle #1}$\hskip3.pt\crcr}}}
\def\otto{{\kern-1.truept\leftarrow\kern-5.truept\to\kern-1.truept}}

\def\mbox{\hbox}
\def\Tr{{\rm Tr}\,}

\def\={{\equiv}}

\def\NoBlackBoxes{\global\overfullrule0pt}

\def\n{\nabla}
\def\supt{\sup_{t\in [0, \bar t]}}
\def\initfiat#1#2#3{
\mgnf=#1
\driver=#2
\tipi=#3
\TIPITOT
\ifnum\mgnf=0
\magnification=\magstep0\hoffset=0.cm
\voffset=-1truecm\hoffset=-.5truecm\hsize=16.5truecm \vsize=25.truecm
\baselineskip=15pt 
\parindent=12pt
\lineskip=4pt\lineskiplimit=0.1pt \parskip=0.1pt plus1pt
\def\ds{\displaystyle}\def\st{\scriptstyle}\def\sst{\scriptscriptstyle}
\font\seven=cmr7
\fi
\ifnum\mgnf=1
\magnification=\magstep1\hoffset=0.cm
\voffset=-1truecm\hoffset=-.5truecm\hsize=16.5truecm \vsize=24.truecm
\baselineskip=15pt 
\parindent=12pt
\lineskip=4pt\lineskiplimit=0.1pt \parskip=0.1pt plus1pt
\def\ds{\displaystyle}\def\st{\scriptstyle}\def\sst{\scriptscriptstyle}
\font\seven=cmr7
\fi
\setbox200\hbox{$\scriptscriptstyle \data $}
}

\def\11{\hbox{l}\!\!\!\hbox{1}\,}

\def\pa{\parallel}
\def\sqr#1#2{{\vcenter{\vbox{\hrule height.#2pt
\hbox{\vrule width.#2pt height #1pt \kern#1pt
\vrule width.#2pt}
\hrule height.#2pt}}}}
\def\square{\mathchoice\sqr56\sqr56\sqr{2.1}3\sqr{1.5}3}
\def\qed{$\square$}

\def\picture #1 by #2 (#3){
\vbox to #2{
\hrule width #1 height 0pt depth 0pt
\vfill
\special{picture #3}
}
}

\def\scaledpicture #1 by #2 (#3 scaled #4){{
\dimen0=#1 \dimen1=#2
\divide\dimen0 by 1000 \multiply\dimen0 by #4
\divide\dimen1 by 1000 \multiply\dimen1 by #4
\picture \dimen0 by \dimen1 (#3 scaled #4)}
}

\def\und(#1){$\underline{\hbox{#1}}$}

\def\xp(#1){\hbox{\rm e}^{#1}}
\input amssym.def

\let\a=\alpha \let\b=\beta  \let\d=\delta
\let\e=\varepsilon
\let\f=\varphi \let\g=\gamma  
\let\l=\lambda

\let\o=\omega  
\let\r=\rho \let\s=\sigma \let\t=\tau 
 \let\x=\xi 
  \let\G=\Gamma \let\L=\Lambda 
\let\O=\Omega

 
\def\mbox{\hbox}
\def\Tr{{\rm Tr}\,}

\def\={{\equiv}}
\def\W#1{#1_{{\kern-3pt\lower6.5pt\hbox{$\widetilde{}$}}%
\kern3pt_{\raise0pt\hbox{\kern0.7pt\ottopunti${}$}}}}

\def\neq{=\hskip-.32cm/\hskip.18cm}
\def\sneq{=\hskip-.18cm/\hskip.07cm}
\def\head (#1,#2){\vskip 1cm
\numsec= #1
\numfor= 1
{\noindent\bf #1. #2.}\vskip .7cm}
\def\theorem(#1,#2){\vskip.3cm\noindent{\bf Theorem #1.\ }{\it #2.}}
\def\lemma(#1,#2){\vskip.3cm\noindent{\bf Lemma #1.\ }{\it #2.}}
\def\propos(#1,#2){\vskip.3cm\noindent{\bf Proposition #1.\ }{\it #2.}}
\def\un#1{\underline{#1}}


\let\pt=\partial

\def\defi{\,{\buildrel def \over =}\,}
\def\sqr#1#2{{\vcenter{\vbox{\hrule height.#2pt
\hbox{\vrule width.#2pt height #1pt \kern#1pt
\vrule width.#2pt}
\hrule height.#2pt}}}}
\def\square{\mathchoice\sqr56\sqr56\sqr{2.1}3\sqr{1.5}3}
\def\qed{$\square$}
\def\v{\varphi}
\def\bstar{\bigcirc\hskip-.34cm *\hskip.2cm}
\def\IR{{\rm I\kern -1.8pt{\rm R}}}
\def\IP{{\rm I\kern -1.6pt{\rm P}}}
\def\IZ{{\rm Z\kern -4.0pt{\rm Z}}}
\def\IC{\ {\rm I\kern -6.0pt{\rm C}}}

\NoBlackBoxes
\initfiat {1}{1}{2}
\magnification=1200
\baselineskip15pt
\centerline{\bf Binary Fluids with Long Range Segregating Interaction
I:}
\centerline{\bf Derivation of Kinetic and Hydrodynamic Equations.}
\vskip.3cm
\centerline{by}
\vskip1cm
\centerline{S. Bastea\footnote{$^\dagger$}
{\eightrm Lawrence Livermore National Laboratory, P. O. Box 808,
Livermore, CA 94550, USA},\hskip.2cm R.
Esposito\footnote
{$^*$}
{\eightrm Dipartimento di Matematica,
Universit\'a degli Studi di L'Aquila, Coppito, 67100 L'Aquila, Italy},
\hskip .2cm
J. L. Lebowitz\footnote
{$^+$}
{\eightrm Departments of Mathematics and
Physics, Rutgers University, New Brunswick, NJ
08903, USA}
\hskip.1cm and \hskip.1cm
R. Marra\footnote
{$^\#$}
{\eightrm Dipartimento di Fisica and Unita INFM, Universit\`a di
Roma Tor Vergata, 00133 Roma, Italy.}
}
\vskip .8cm
\bigskip
\bigskip
{\baselineskip = 10pt\rightskip1.4cm\leftskip 1.4cm \ottopunti

{\noindent {\bf Abstract:}\/}
We study the evolution of a two component fluid consisting of ``blue''
and ``red''
particles which interact via strong short range (hard core) and weak
long range pair
potentials. At low temperatures the equilibrium state of the system is
one in which
there are two coexisting phases. Under suitable choices of space-time
scalings and
system
parameters we first
obtain (formally) a mesoscopic kinetic
Vlasov-Boltzmann
equation for the one
particle position and velocity distribution functions, appropriate for
a
description of the phase segregation kinetics in this system. Further
scalings
then yield
Vlasov-Euler and
incompressible Vlasov-Navier-Stokes equations. We also obtain, via the
usual truncation of
the Chapman-Enskog
expansion, compressible Vlasov-Navier-Stokes equations.
\bigskip
\bigskip
\bigskip}
\baselineskip=18pt
\head(1,Introduction)

The process of phase segregation in which a system evolves from an
initial unstable
homogeneous state into a
final equilibrium state consisting of two coexisting phases is of
continuing
theoretical and practical
interest [GSS], [FLP], [L]. Such a process occurs whenever the system,
which
is initially at
values of the
thermodynamic parameters, say temperature $T_0$ and pressure $p_0$,
corresponding to a
single homogeneous phase
has its parameters ``suddenly'' changed to new values, say $T$ and $p$,
at which there
is a coexistence of
phases.

This happens, for example, when an alloy is `quenched' from a
high temperature
melt or solid to a low
temperature
solid state by sudden cooling [GSS]. After such a quench the system
finds itself in
an unstable (or metastable)
situation, as far as the spatial concentrations, which have not been
able to adjust
rapidly enough to the "sudden"
quench, are concerned and domains
of the two equilibrium phases form and start growing in time. This
proceeds until
there are ``in the
final state'' only two
regions of pure
equilibrium phases separated by an interface. Since the kinetics of the
domain growth
have a profound influence
on the
properties of the alloy, this problem has been and continues to be
extensively studied
both theoretically and
experimentally
[GSS]. For such alloy systems the segregation process takes place mainly
through the
(anti)diffusion of the two
components---from a uniformly mixed state to a demixed one. There are
no
macroscopic matter or energy flows
since the system
is a solid and has a high heat conductivity which keeps the
temperature equal to
some constant ambient value.
The
only relevant conserved quantities are therefore the particle numbers of
the two
components and the macroscopic
equations describing
the process are fairly well established: these are the well known
Cahn-Hilliard
equations [CH] and variations on
them. We refer
the reader to reviews on this subject [GSS].

The situation is much less clear for phase segregation in fluids where
macroscopic
flows of matter and heat
are important. There are now additional conservation laws for momentum
and energy and
there is no general
consensus even on what hydrodynamical equations are most appropriate
for describing
the macroscopic
evolution of the system [S], [OP], [AGA]. In particular, it is not clear
which is the
correct
coupling between the
Cahn-Hilliard equation for the order parameter and the Navier-Stokes
equation for the
fluid velocity.

To make a start on the mathematical
analysis of such processes we investigate a model binary fluid
introduced in [BL]
where the process of phase
segregation was studied numerically.

In the present work we derive general equations appropriate both in the
one phase and in the
coexistence region. In part II we consider applications to the
segregation process
including an analysis of new numerical results. Many of our discussions
here will be
semi-heuristic. In particular, we will not go into detail about the
domain
of validity of the
technical conditions necessary for the rigorous mathematical
establishment of the results.

The model we study is composed of two types of particles, call them red
and blue. There are $N_r$
red and $N_b$ blue particles in a cubic box of volume $\Lambda=L^d$; we
will generally consider $d=3$ and use periodic boundary conditions.
The particles all have unit mass and hard core diameter $a$.
Particles of different kind
also interact with each other through a long range pair potential of the
Kac type, having a
range $\ell$ and a strength $A_\ell$. By properly choosing $A_\ell$, we
obtain, in the
limit $\ell\to\infty$, a system whose equilibrium properties are
described by a mean-field
type phase diagram exhibiting a demixing phase transition for
temperature $T<T_c$ [LP].

This transition is essentially independent of the hard core size $a$
and
the
dimensionless microscopic particle
densities $\rho_r a^3$ and $\rho_b a^3$ can therefore be arbitrarily
small
in the demixed phases. This means that we can have a situation in
which,
at least in principle,
the whole phase transition is well described by a Vlasov-Boltzmann type
of kinetic
equation. We will in fact
see that we can, by suitably scaling space and time and the densities,
obtain, at least on the formal level, a set of nonlinear
Vlasov-Boltzmann (VB)
equations, describing the evolution
of the one
particle distribution functions $f^\a(q,v,t)$, $\a=r,b$.

The VB equations we derive are of a form similar to ones conjectured
for a one
component fluid with hard
cores and an attractive long range interaction [DS], [G].
Such a system however requires the hard cores for
stabilization against collapse [LP] and $\rho a^3$ is greater than $1/3$
in the liquid
phase.
It is therefore not clear that a VB equation is an appropriate kinetic
description of
such a liquid-vapor transition. This is the motivation for introducing
the binary model we
consider here.

We discuss the scalings necessary to go from a microscopic Hamiltonian
description of
the time evolution to the
VB equations in Section 2 leaving a formal derivation, in the spirit of
Cercignani [C] and Lanford [Lan], to
Appendix C. The equations
themselves are of the same form as those used in [BL] for the kinetics
in the
coexistence regime
of this system. Their numerical results for the time evolution and the
analysis of the
stationary states showed
that these
VB equations for the one particle distributions $f^\a$ indeed lead to
the phase
segregated state expected from
purely equilibrium considerations.

While the mesoscopic description in terms of the one-particle
distribution
functions is a great
simplification compared to
the full microscopic representation, it is still more complicated than
the macroscopic
theory that treats the binary
system as a continuum with well defined local density
$\rho(x,t)$,
concentration difference $\varphi(x,t)$,
velocity $u(x,t)$ and temperature $T(x,t)$.
The derivation of hydrodynamic equations from the Boltzmann equation
(which one expects to be structurally of the same form as those
describing dense binary
fluids) is closely
related to the problem of
finding
approximate solutions of the Boltzmann equation. The reason for
this is that the fluid dynamic variables are defined and change
on space and time
scales which are very large when measured in units of the mean free
path
and mean free
time between collisions,
i.e. the kinetic
or mesoscopic scale. Therefore, it can be expected that the system
will reach a
state close to local equilibrium in a macroscopically very small time
interval,
meaning that $f^\a(x,v,t)$ should
stay close to local Maxwellians, with parameters
$\rho^\a$,
$ u$ and
$ T$, which change slowly on the kinetic scale. The big disparity
between the kinetic
and hydrodynamic scales
suggests
looking for a solution of the Boltzmann equation as a series expansion
in the scale
parameter which is the
ratio of these two scales. Many rigorous results in this direction
have been
obtained in recent years,
especially for the Euler (E) and the incompressible Navier-Stokes
(INS) equations.
The situation is
less satisfactory in the case of the compressible Navier-Stokes (NS).
This is a
consequence of the fact that
while the E
and INS equations correspond to well defined scaling limits, in which
the mean free
path goes to zero, there
is no such
scaling limit for the NS equations
as can be seen from the fact that these equations are not invariant
under scaling [DEL].

Having obtained the VB
equations we
turn to the derivation of hydrodynamic equations.
The results available for these equations are fewer than for the
Boltzmann
equation. In
Sect. 3 we present a rigorous derivation of the Vlasov-Euler (VE)
equations for this
system, which differs from
the
usual Euler equations by the presence of self-consistent forces coming
from
the Vlasov
terms. We do this by adapting
to
this case the method of Caflisch [Ca80], i.e. we prove that
the Hilbert expansion is asymptotic, by showing that the remainder at
any order is
finite in a suitable
Sobolev norm.

We then consider in Sect. 4 and 5 a modified
Chapman-Enskog
expansion of the kind
considered by Caflisch [Ca87] and show also in this case that the
remainder at any
order is finite in the
same Sobolev norm. The term of zero order in this expansion is a
Maxwellian with
parameters solving a
set of dissipative new PDE's, the Vlasov-Navier-Stokes (VNS) equations,
where, beyond
the usual terms present
in the
compressible Navier-Stokes equations, there are diffusive terms coming
from the
presence of the
self-consistent force. In particular, the equation for the
concentration can be put
in the form of a
gradient flux of an energy functional [BELMII] which is similar to an
exact
evolution equation
derived for a microscopic
model of a binary alloy. The latter has been proven to yield the same
late
time phase
segregation behavior as the
Cahn-Hilliard
equation, [GL96], [GL97]. Both Vlasov-Euler and Vlasov-Navier-Stokes
have non trivial
stationary solutions
with the same solitonic profile as in the BV equation.

Finally in Sect. 6 we consider the incompressible
regime for these equations and derive, under suitable initial conditions
and scaling,
a set of PDE's with
dissipative terms
involving a force linear in the concentration (they are essentially the
linearization
of the analogous terms in
the
compressible equations around a constant concentration and density
profile).
Above results all rely on the crucial assumption that the initial value
problems for
the hydrodynamical equations
have a unique smooth solution at least on some macroscopic time
interval. We do not
discuss the technical
conditions which ensure the existence of such solutions.

\vskip.2cm
\numsec= 2
\numfor= 1
\head(2, Vlasov-Boltzmann equation for a binary mixture)

We consider a system of $N_r$ {\it red} particles with positions
$\x_i^r$ and velocities $v_i^r$, $i=1,\dots,N_r$ and
$N_b$ {\it blue} particles with positions
$\x_i^b$ and velocities $v_i^b$, $i=1,\dots,N_b$, in a $3$-dimensional
torus $\L$,
interacting
via two body forces. $N=N_r+N_b$ is the total number of particles. The
potential
energy is

$$\eqalign{&V(\x^r_1,\dots,\x^r_{N_r};\x^b_1,\dots,\x^b_{N_b})= {1\over
2}A_\ell\sum_{\a\sneq\b} \sum_{i=1}^
{N_\a}
\sum_{j=1}^ {N_\b} U_\ell(|\x^\a_{i}-\xi_{j}^\b|)\cr& + {1\over
2}\sum_{\a,\b} \sum_{i=1}^
{N_\a} \sum_{j=1}^
{N_\b}W_a(|\x^\a_{i}-\x_{j}^\b|)}
\Eq(0.1)$$ where $\a,\b=r,b$, $U_\ell$ is the long range potential
$$ \ U_\ell(r)=U( {r\over \ell})
\Eq(o.2)$$
for some bounded, smooth non-negative function $U$ on $\IR_+$. The
factor $A_\ell$ is the intensity of the long range interaction to be
suitably
chosen to get a mean field type of behavior when $\ell$ becomes very
large
compared to the interparticle spacing [LP]. The potential
$W_a$ is the {\it formal} hard core potential

$$ {\cal W}_a(r) = \cases{ \infty \quad &\hbox{ if $r<a$
}\cr
0 \quad &\hbox{ otherwise.}}$$
\vskip.1cm
In other words the particles are hard spheres of diameter $a$
interacting
by elastic collisions which are color blind and by a weak repulsive long
range force
between
particles of different species. The total
number of particles of each species as well as the total momentum and
energy are invariant
during the evolution.

Choosing the size of $\Lambda$ to be $\ell$ (or some constant multiple
thereof)
there are two characteristic length scales for
this dynamics:
$a$, the range of the hard core potential and
$\ell$, the range of the Kac potential. We can consider a third length,
which depends on the density,
the
mean free path $\lambda$ defined by the relation
$$\l= {\ell^3\over Na^2}.$$
The kinetic limit arises when there is a large separation between
$a$ and
$\lambda$, corresponding to a low density $(N/\ell^3)$ situation.
To obtain a kinetic limit we send $N$ and $\ell$ to
$\infty$ while $a$ is fixed, say $1$, in such a way that
$\l/ \ell$ is finite ($N\sim \ell^2$) 
and assume initial data almost constant on regions of size $\sim\l$.
We denote by
$$\d={a\over \l}={1\over \l}$$
and assume finite
$$\g={\l\over \ell}$$
The kinetic equations will be obtained in the limit $\d\to 0$, assuming
$$A_\ell= \gamma^3 \d^2,$$
meaning that $A_\ell$ is proportional to $1/N$. A further
limit $\g\to
0$ will provide the
hydrodynamical limit to be discussed later.

In kinetic coordinates $q$, that is
$q_i^\a=\d\xi_i^\a$, for
$\a=r,b$ and $i=1,\dots,N_\a$ and kinetic time
$\tau=\d \tau_m$, $\tau_m$ being the microscopic time, the equations of
motion for
the system
are, for $\a=r,b$ and $i_\a=1,\cdots,N_\a$

$$\eqalign{
{d q_{i_\a}^\a\over d \tau}&= v_{i_\a}^\a\cr
{d v_{i_\a}^\a\over d\tau}&=
\g^3\d^{2}\sum_{j_\b=1}^{N_\b}K(\g|q_{i_\a}^\a-q_{j_\b}^\b\
|)(1-\d_{\a\b})} \Eq(?)$$
in $$\G_N=\{(q_1,v_1,\dots, q_N,v_n)\in \L^N\times \IR^{3N}\,| \,\,
|q_{i}-q_{j}|>\d
,\quad
i\neq j\},$$
where $K(|x-y|)=-(\nabla U)(|x-y|)$, $N=N_r+N_b$, and we use the
notation $q_i, v_i$,
$i=1,\dots,N$ when the color is irrelevant. When two particles are in
contact (namely
at distance $\d$) they
undergo an elastic collision regardless of their color. We neglect the
event that more
that two particles
are in contact because it has vanishing Lebesgue measure. Hence the
evolution \equ(?)
is defined
only almost everywhere.

Note that when
$N\sim \ell^2$ then the mean force on each particle on the kinetic
scale \equ(?) is of order unity.
This is the reason why our original choice of the strength
$A_{\ell}$ of the
potential in \equ(o.2) was like $\ell^{-2}$
rather than $\ell^{-3}$ as in the usual case [LP].

With this scaling we can get, at least formally, in the limit $\d\to 0$
the
Vlasov-Boltzmann equation for a
binary
mixture of hard core particles interacting via a weak long range
potential. A formal
proof of this is given in
Appendix C. The rigorous proof would require the extension of the
Lanford argument to
this case, an extension
that
is not obvious because the Vlasov part is not well controlled in the
Lanford norms.

Even if we have discussed the derivation of the Vlasov-Boltzmann
equation only for
hard spheres, from now on we
consider the Vlasov-Boltzmann equations in full generality. The
function $f^{r}(q,v,\t)$ (resp. $f^{b}(q,v,\t)$) is proportional to the
probability
density of finding a red
(resp. blue) particle at $q\in
\O\subset \IR^3$, with velocity $v\in \IR^3$ at time $\t\ge 0$.
We notice that the relation between the $f^\a$'s and the microscopic one
particle
densities
$\rho_{1}^\a (\xi, v, \tau_m)$ (normalized to $N_\a$)
is given by
$$f^\a (q, v, \tau) =\lim_{\d\to 0}\delta^{-1}\rho_{1}^\a (\d^{-1} q, v,
\d^{-1}\tau).$$

The functions $f^{r}$ and $f^{b}$ are positive and normalized to
$\gamma^{-3}$ for any
value of $\t$. They are
solutions to the equations
$$\eqalign{&\pt_\t f^{r} +v\cdot \n_qf^{r} +F^{r}\cdot \n_v f^{r} =
J(f^{r},f^{r})+J(f^{r},f^{b}),\cr&
\pt_\t f^{b} +v\cdot \n_qf^{b} +F^{b}\cdot \n_v f^{b} =
J(f^{b},f^{b})+J(f^{b},f^{r}).}
\Eq(2.1)
$$
The Vlasov force acting on each particle is of the Kac type, meaning
that for any
$\g>0$,
the forces are conservative non local forces with range $\g^{-1}$
defined by the
position
$$F^{\a}(q,\t)= -\n_q\int_\O d\/q' \g^3 U(\g|q-q'|)n^{\b}(q',\t),\quad
\a=r,b,
\quad\a\ne \b\Eq(2.2)$$
with $U(|q|)$ a smooth, non negative function of compact support and
$n^r$, $n^b$ are
the rescaled spatial densities of the red and blue particles:
$$ n^{\a}(q,\t)=\int_{\IR^3} d\/v f^{(\a)}(q,v,\t),\qquad
\int_{\Omega} d\/q\ n^{(\a)}(q,\t)=\gamma^{-3}.
\Eq(2.4)
$$
For any positive functions $f$ and $g$, $J(f,g)$ denotes the effect of
the collisions
of
particles distributed according to $g$ on the distribution $f$. Its
expression is
given
by
$$J(f,g)=\int_{\IR^3} d\/v_*\int_{S^2}d\/\omega b(|v-v_*|,\omega)
[f(v')g(v'_*)-f(v)g(v_*)].
\Eq(2.3)$$
Here $b(|v|,\o)$ is the differential cross section of the short range
interaction,
$\o\in S_2$ is the impact parameter and $v',v'_*$ are
the incoming velocities corresponding to an elastic collision with
outgoing
velocities $v,v_*$ and impact parameter $\o$.
We assume the Grad's (see [Gra]) angular cutoff condition that
$b(|v|,\o)$ is a smooth function growing at most linearly for large
$|v|$, i.e.
$b(|v|,\o)=|v|^\s h(\o)$ with
$0\le \s\le 1$ and $h$ a smooth bounded function on $S_2$.
\vskip.2cm
An important property of the collisions is the entropy production
inequality:
let
$${\cal N}_\a=\int_{\IR^3} dv J(f^\a,f^\a)\log f^\a, \quad \a=1,2,$$
$${\cal N}_{\a,\b}=\int_{\IR^3} dv J(f^\a,f^\b)\log f^\a, \quad
\a,\b=1,2.$$
Then ${\cal N}_\a$'s as well as ${\cal N}_{1,2}+{\cal N}_{2,1}$ are non
negative. Moreover
$N_\a$ vanishes
as usual if and only if the $f^\a$'s are Maxwellians:
$$f^\a= M(n^\a,u^\a,T^\a;v),\quad \a=1,2$$
with
$$M(n,u,T;v):={n\over(2\pi T)^{3/2}}{\displaystyle\hbox{\rm
e}^{-(v-u)^2/2T}}.\Eq(maxw)$$
Furthermore ${\cal N}_{1,2}+{\cal N}_{2,1}$ vanishes if and only if the
two
Maxwellians have the
same local temperature and mean velocities:
$$u^\a=u, \quad T^\a=T, \quad \a=1,2.$$
This implies that the
only solutions of the equations
$$\eqalign{&J(f_1,f_1)+J(f_1,f_2)=0, \cr&J(f_2,f_2)+J(f_2,f_1)=0}$$
are Maxwellians with the same mean velocity and temperature. General
arguments suggest that
all the stationary solutions of equations \equ(2.1) will be Maxwellians
with $u=0$, T(q)=T, and
densities satisfying the equations
$$
\eqalign{
T\log n^1(q)&+\int dq' \gamma^3U(\gamma |q-q'|) n^2(q')=C_1\cr
T\log n^2(q)&+\int dq' \gamma^3U(\gamma |q-q'|) n^1(q')=C_2.
}
\Eq(1.1)$$

Beyond the spatially constant equilibria, there may be other
spatially non
homogeneous solutions.
For example, by prescribing the boundary conditions in one dimension
$$\lim_{z\to \pm \infty}n^\a(z)= \bar n^\a_{\pm},$$
one gets at small values of $T$ a solitonic solution describing the
interface profile [BL].
We shall leave a discussion of this part for [BELMII] and focus here on
deriving
macroscopic equations for the evolution of the conserved quantities.

Before closing this section, let us define
$$f(q,v,\t)= {1\over 2}[f^r(q,v,\t)+f^b(q,v,\t)]$$
as the density of finding a particle at $q$ with
velocity $v$ at time
$\t$, independently of its color.
Moreover, we set
$$\phi(q,v,\t)= {1\over 2}[f^r(q,v,\t)-f^b(q,v,\t)].$$
The system \equ(2.1) can be written
in the following equivalent form:
$$\eqalign{&\pt_\t f +v\cdot \n_q f +2F\cdot \n_v f+ 2W\cdot \n_v\phi =
4J(f,f),
\cr&\pt_\t \phi +v\cdot \n_q \phi +2F\cdot \n_v \phi+ 2W\cdot \n_vf =
4J(\phi,f),}
\Eq(2.5)$$
where $F=F^{r}+F^{b}$, $W=F^{r}-F^{b}$.
We can absorb the numerical factors by redefining $U$ as $U/2$ and $b$
as $b/4$
so obtaining
$$\eqalign{&\pt_\t f +v\cdot \n_q f +F\cdot \n_v f+ W\cdot \n_v\phi =
J(f,f),
\cr&\pt_\t \phi +v\cdot \n_q \phi +F\cdot \n_v \phi+ W\cdot \n_vf =
J(\phi,f).}
\Eq(2.5')$$
\vskip.3cm
\numsec=3
\numfor=1
\head(3, Compressible Hydrodynamics)
We are interested in the behavior of the system on the macroscopic
scale. To this end
we
introduce a scaling parameter $\e$ representing the ratio between the
kinetic and
macroscopic space units and, for any $t\ge 0$ and $x\in \e\O$ we set
$$\t=\e^{-1}t,\quad q= \e^{-1}x.$$
We assume that at time zero the densities vary slowly on
the microscopic scale $ f^i(q,v,0)=\tilde f(\e q, v, 0)$ and look for
solutions of \equ(2.5') such that
$$f^i(q,v,\t)=\tilde f^i(\e q, v, \e \t),\quad i=r,b$$
with $\tilde f^i$ smooth functions on $\e\O\times \IR^3\times \IR_+$.
For the force we have
$$F^r(\e^{-1}x,\e^{-1}t)=-\e\n_x\int_{\e \O}dx'\left({\g\over
\e}\right)^d
U\left[\left({\g\over \e}\right)|x-x'|\right]\tilde n^b(x',t),$$
and a similar relation for $F^b$.
Therefore, if we assume $\g=\e$, also the forces are slowly varying
functions and the
$\tilde f$ satisfy the following system, where we remove the ``tilde's''
because in
the
sequel we shall always consider only the macroscopic variables:
$$\eqalign{&\pt_t f +v\cdot \n_x f +F\cdot \n_v f+ W\cdot \n_v\phi =
\e^{-1}J(f,f),\cr&
\pt_t \phi +v\cdot \n_x \phi +F\cdot \n_v \phi+ W\cdot
\n_vf = \e^{-1}J(\phi,f).}
\Eq(2.7)$$
We shall use the notation
$$F={\bf K}\bstar f, \quad W=-{\bf K}\bstar \phi,\Eq(2.8)$$
where,
$${\bf K}(x)=-\n_x U(|x|)\Eq(2.8.0)$$
and, for any function $g$ we set
$$({\bf K}\bstar g)(x,t)\defi \int_\O dx'{\bf K}(|x-x'|)\int_{\IR^3} dv
g(x',v,t).\Eq(2.8.1)$$

We will show that the solution of the system \equ(2.7) is close for
$\e$ small to the local
equilibrium with parameters $\r^{(1)}$, $\r^{(2)}$, $u$ and $T$
satisfying the following set
of hydrodynamic equations:
$$\eqalign{&\pt_t \r +\n\cdot [\r u]=0,\cr&
\partial_t \varphi + \nabla\cdot(\varphi u) =\e
\nabla\cdot(D Q),
\cr&\rho D_t u +\nabla P -{\rho\bf K}*\rho + \varphi{\bf K}*\varphi=
-\e\nabla \sigma,
\cr&{3\over 2} \rho D_t T +P \nabla\cdot u=\e\nabla(\kappa\nabla T)
-\e\sigma: \nabla
u -\e{\bf K}*\f\cdot
D Q.}
\Eq(4.11)$$
Here $\r=\r^{(1)}+\r^{(2)}$ is the total density,
$\f=\r^{(1)}-\r^{(2)}$, $P=\rho T$
$$D_t:=\pt_t + u\cdot \n,$$
$$\sigma:=-\nu(\n u+\n u^\dagger-{2\over
3}\Bbb I\,\n\cdot
u)$$
$$Q:=\nabla{\v\over\rho} + {1\over \rho^2T} (\rho^2-\f ^2){\bf
K}*\f.\Eq(4.11.1)$$
$\n u^\dagger$ is the adjoint of the matrix $\n u$, $\s:\n u= \Tr (\s\n
u)$, $\Bbb I$
is the unit matrix, $\nu$ and $D/\rho$
are the viscosity and the diffusion coefficients and
$\kappa$ is the heat conductivity. These are computed from the VBE. The
above equations, with $\e=0$ will be referred to as the Vlasov-Euler
equations (VE). We assume that the initial value problem for such
equations, with suitable
initial data, has a sufficiently smooth solution at least on a time
interval $[0,\bar t]$.
Under such conditions we will prove in the next section and in Appendix
A that the solution to the
VBE for the binary fluid, under the Euler scaling, converges to the
Maxwellian
local equilibrium with parameters satisfying the VE equations in the
interval $[0,\bar t]$,
with an error of order
$\e$ (Proposition 4.1 and Corollary 4.2).

When $\e>0$, the above equations will be referred to as the
Vlasov-Navier-Stokes equations (VNS).
In Section 5, using also the arguments of Appendix A we will show that
their solutions
provide an approximation up to the order $\e^2$ to the solutions of the
VBE in
the Euler scaling, provided that the initial value problem for such
equations has suitably
smooth solutions as before. The precise statement is given in
Proposition 5.1 and Corollary
5.2.

Like for the usual Navier-Stokes equations, which are frequently and
successfully used with
$\e=1$ in physical and engineering applications, although their
derivation is
restricted to small values of
$\e$, we will consider the VNS equations with
$\e=1$ and analyze some of their properties in [BELMII].
In order to get diffusive effects as sharp limits of the VBE, it is
necessary to go to the parabolic scaling where $\tau=\e^{-2}t$ and
consider
simultaneously a low Mach number situation. This will be discussed in
Section 6.

\vskip.8cm
\numsec=4
\numfor=1
\head(4, Euler limit)
We outline the proof of the convergence of the Vlasov-Boltzmann system
to the VE equations.
The proof will be completed in Appendix A. We fix the Maxwellian
$M(\r,u,t)$ with $\r,u,T$
possibly depending on space and time and
denote
$${\cal L} f= J(M,f)+J(f,M)
\Eq(2.9)$$
and
$$\Gamma f=J(f,M).\Eq(2.9.5)$$
Moreover, we set
$$Q(f,g)={1\over 2}[J(g,f)+J(f,g)]
\Eq(2.10)$$
It is easy to check that, as for the one-component Boltzmann equation,
$${\cal L} f=0\quad \hbox{ iff } \quad f=M\chi_\a,\quad \a=0,\dots,4,
\Eq(2.9.7)$$ where
$$\chi_0=1,\quad
\chi_i=v_i,\quad i=1,\dots,3,\quad \chi_4=v^2/2
\Eq(2.11).$$
Moreover, along the same lines one gets
$$\Gamma f=0 \hbox{ iff } \quad f=aM, \quad a\in \IR.\Eq(2.11.2)$$

We shall try to solve \equ(2.7) following [Ca80], in terms of a
truncated
Hilbert
expansion of the form
$$\eqalign{&f=\sum_{n=0}^K\e^n f_n+\e^m R_f,\cr&\phi=\sum_{n=0}^{K}\e^n
\phi_n+\e^m R_\phi,}\Eq(2.12.5)$$
with suitably chosen positive integers $K$ and $m$.
The
functions
$f_n$ and $\phi_n$ are computed using a Hilbert expansion and the
remainders $R_f$ and $R_\phi$ are defined as the difference between the
solution and the truncated expansion.
\vskip.5cm

We substitute in \equ(2.7) the formal power series
$$f=\sum_{n=0}^\infty \e^n f_n, \quad \phi=\sum_{n=0}^\infty \e^n
\phi_n,\Eq(3.1)$$
$$F=\sum_{n=0}^\infty \e^n F_n=\sum_{n=0}^\infty \e^n {\bf K}\bstar
f_n,\quad
W=\sum_{n=0}^\infty \e^n W_n=\sum_{n=0}^\infty \e^n {\bf K}\bstar
\phi_n,\Eq(3.2)$$
and denote by $D_t$ the time derivative along the trajectories:
$$D_t=\pt_t+v\cdot\n_x .$$
We have:
$$\e^{-1} Q(f_0,f_0)+\sum_{n=0}^\infty\e^n\Big[
2Q(f_0,f_{n+1})+{\cal S}_n\Big]=0,\Eq(3.4)$$
$$\e^{-1} J(\phi_0,f_0)+\sum_{n=0}^\infty\e^n\Big[
J(\phi_{n+1},f_0)+ J(\phi_0,f_{n+1})+{\cal T}_n\Big]=0,\Eq(3.5)$$
where
$${\cal S}_n=\sum_{\st (h,h'):\ h, h'\ge 1
\atop \st h+h'=n+1}Q(f_h,f_{h'})-\sum_{\st (h,h'):\ h,h'\ge 0\atop
h+h'=n}
\big[F_h\cdot \n_v f_{h'} +W_h \cdot \n_v\phi_{h'}\big]-D_t
f_{n}.\Eq(3.6)$$
$${\cal T}_n=\sum_{\st (h,h'):\ h, h'\ge 1
\atop \st h+h'=n+1}J(\phi_h,f_{h'})-\sum_{\st (h,h'):\ h,h'\ge 0\atop
h+h'=n}
\big[F_h\cdot \n_v \phi_{h'} +W_h \cdot \n_vf_{h'}\big]-D_t
\phi_{n}.\Eq(3.7)$$
In order that the formal series solve \equ(2.7) the coefficients have to
satisfy the conditions:
$$\eqalign{&Q(f_0,f_0)=0,\cr& J(\phi_0,f_0)=0}\Eq(3.8)$$
and, for any $n\ge 0$,
$$\eqalign{&2Q(f_0,f_{n+1})+{\cal S}_n=0,\cr& J(\phi_{n+1},f_0)+
J(\phi_0,f_{n+1})+{\cal T}_n=0.}\Eq(3.9)$$
As remarked in the previous section, the first of the conditions
\equ(3.8), implies
that
$f_0$ is a Maxwellian with parameters depending on $x,t$:
$$f_0(x,v,t)=M(\r(x,t),u(x,t),T(x,t);v):=M(v).\Eq(3.10)$$
Moreover from the second eqn. of \equ(3.8) we get
$$\phi_0(x,v,t)={\f(x,t)\over \r(x,t)}M(\r(x,t),u(x,t),T(x,t);v),$$
for some suitable function $\f(x,t)$.
Using \equ(2.9) and \equ(2.9.5) we can write \equ(3.9) as
$$\eqalign{&{\cal L}f_{n+1}=-{\cal S}_n,\cr& \G\phi_{n+1}=-
J(\phi_0,f_{n+1})-{\cal T}_n.}\Eq(3.11)$$
Since ${\cal S}_n$ and ${\cal T}_n$ only depend on the $f_k$ and
$\phi_k$ for $k\le
n$, we
have first to solve the first eqn. and then, once $f_{n+1}$ is
determined, we
solve the second one for $\phi_{n+1}$.

In order to check the solvability of these equations we introduce the
Hilbert
space ${\cal H}$ of measurable functions on $\IR^3$ such that the
scalar product
$$(f,g)=\int_{\IR^3}dv f(v) g(v) M^{-1}(v),\Eq(3.12)$$
is finite. In this Hilbert space the
operators
${\cal L}$ and $\G$ are densely defined and symmetric. Moreover, the
null
spaces of ${\cal L}$ and $\G$ are the five-dimensional subspace spanned
by
$\{M\chi_\a, \a=0,\dots,4\}$ introduced in \equ(2.11) and the
one-dimensional
space spanned by
$M\chi_0$ respectively. We denote by ${\cal P}$ and ${\cal K}$ the
projectors
on such subspaces and by ${\cal P}^\perp=1-{\cal P}$ and ${\cal
K}^\perp=1-{\cal K}$ the projectors on their orthogonal complements.
 From the properties of ${\cal L}$ and $\G$ it is immediate to check that
$${\cal P}{\cal L}=0, \quad {\cal K}\G=0.\Eq(3.13)$$

Both ${\cal L}$ and $\G$ are non positive and there are positive
constants $\d$ and $\d'$
$$(f,{\cal L}f)\le -\d\|{\cal P}^\perp f\|^2,\quad (f,{\G}f)\le
-\d'\|{\cal
K}^\perp f\|^2.\Eq(3.14)$$
Moreover
$$\eqalign{&{\cal L}=-\nu + K,\cr& \G= -\nu+\Theta,}\Eq(3.15)$$
where
$$\nu(x,v,t)=\int_{Re^3}dv_*
b(|v-v_*|)M(\r(x,t),u(x,t),T(x,t);v_*),\Eq(3.16)$$
is a strictly positive function such that
$$\nu_0(1+|v|)^\s\le \nu(x,v,t)\le \nu_1(1+|v|)^\s\Eq(3.17)$$
for some positive constants $\nu_0$ and $\nu_1$ provided that $\r$ and
$T$ are
bigger than some fixed positive constants.
Furthermore, $K$ and $\Theta$ are compact operators on ${\cal H}$.
Therefore,
using the Fredholm alternative theorem we can conclude the existence of
solutions
to
\equ(3.11) provided that ${\cal S}_n\in{\cal P}^\perp{\cal H} $ and
${\cal
T}_n+J(\phi_0,f_{n+1})\in{\cal K}^\perp{\cal H}$.
These conditions can be verified inductively, as in the usual
one-component
Boltzmann equation. We now write the conditions for $n=0$ which
determine the
macroscopic equations for $\r$, $u$, $T$ and $\f$:
Since ${\cal P}[\n_v f_0 ]$ and ${\cal P}[\n_v\phi_0]$ have no
component along $\chi_0$, it is easy to check that the condition
${\cal P}{\cal {\cal S}}_0=0$ can be written explicitly as
$$\eqalign{&\pt_t\r +\n_x\cdot[\r u]=0,\cr&
\r[\pt_t u+(u\cdot \n_x)u]=-\n_x P +\r {\bf K}*\r - \f{\bf K}*\f,\cr&
\r[\pt_t e+(u\cdot \n_x)e] + P\n_x \cdot u=0},\Eq(3.18)$$
where $*$ denotes the usual convolution, $P=\r T$ is the equation of
state
for the pressure in the perfect gas and $e=3T/2$ is its internal kinetic
energy.
On the other hand, since ${\cal K}J=0$ and ${\cal K}\n_v=0$, the
condition
${\cal K} {\cal T}_0=0$ becomes ${\cal K}D_t\phi_0=0$ which is
explicitly
written as
$$\pt_t \f+\n_x\cdot[\f u]=0,\Eq(3.19)$$

Equations \equ(3.18) and \equ(3.19) represent the Euler equations for
the binary
mixture. They differ from the usual Euler equations by the presence of
the
equation \equ(3.19) for $\f$ and for the nonlinear self consistent force
terms due to the long range Kac interaction. We will refer to them as
the {\it Vlasov-Euler equations (VE)}. Existence of solutions to the
initial
value problem for the system
\equ(3.18)-\equ(3.19) requires some analysis but we do not discuss this.
We simply assume
that, for sufficiently smooth initial data a unique
solution of the system exists and stays smooth up to some time $\bar
t$.

Given such a solution, the functions $f_1$ and $\phi_1$ can be found by
solving
\equ(3.11) with $n=0$. In consequence, $f_1$ is determined up to $p_1\in
{\cal
P}{\cal H}$ and $\phi_1$ up to $q_1\in{\cal K}{\cal H}$. The procedure
can then continue by taking advantage of the arbitrariness of $p_1$ and
$q_1$ to
satisfy the conditions ${\cal P}{\cal {\cal S}}_1=0$, ${\cal K}{\cal
T}_1=0$. In this way the
functions $f_n$ and $\phi_n$ can be found for any $n$. Classical results
by
Grad [Gra] provide the smoothness and decay properties we use below.

Now we go back to the truncated expansions \equ(2.12.5). Once $\f_n$ and
$\phi_n$ are computed for $n=0,\dots, K$, we can look for the equations
for
the remainders $R_f$ and $R_\phi$. A straightforward calculation shows
that, in
order that $f$ and $\phi$ satisfy \equ(2.7), $R_f$ and $R_\phi$ have to
solve
the equations
$$\eqalign{ D_t R_f + F\cdot\n_v R_f+W\cdot \n_v R_\phi=&\e^{-1} {\cal
L}R_f +{\cal L}^{(1)}R_f+\e^{m-1}[J(R_f,R_f)+A_f],\cr
D_t R_\phi + F\cdot\n_v R_\phi+W\cdot \n_v R_f=&\e^{-1} \G
R_\phi+\e^{-1}\tilde\Theta R_f +\G^{(1)}R_\phi+
\cr&\e^{m-1}[J(R_\phi,R_f)+A_\phi],}\Eq(3.20)$$
where
$$\eqalign{&{\cal L}^{(1)} g = \sum_{h=1}^K
\e^{h-1}[J(f_h,g)+J(g,f_h)],\cr&
\tilde\Theta g= J(\sum_{n=0}^K\e^n \phi_n, g),\qquad
\G^{(1)} g = J(g,\sum_{h=1}^K \e^{h-1}f_h),}\Eq(3.21)$$
$$\eqalign{&A_f=\e^{K-2m+1}\Big(
\sum_{\st (h,h'):\ h, h'\ge 1
\atop \st h+h'> K+1}\e^{h+h'-K-1}
Q(f_h,f_{h'})\cr&-\sum_{\st (h,h'):\ h,h'\ge
0\atop h+h'> K}\e^{h+h'-K}
\big[F_h\cdot \n_v f_{h'} +W_h \cdot \n_v\phi_{h'}\big]-D_t
f_{K}\Big)\cr&
A_\phi=\e^{K-2m+1}\Big(
\sum_{\st (h,h'):\ h, h'\ge 1
\atop \st h+h'> K+1}\e^{h+h'-K-1}
J(\phi_h,f_{h'})\cr&-\sum_{\st (h,h'):\ h,h'\ge
0\atop h+h'> K}\e^{h+h'-K}
\big[W_h\cdot \n_v f_{h'} +F_h \cdot \n_v\phi_{h'}\big]-D_t
\phi_{K}\Big),}\Eq(3.21.1)
$$
and
$$\eqalign{&F=\sum_{n=0}^K\e^n F_n+ \e^m{\bf K}\bstar R_f,
\cr& W= \sum_{n=0}^K \e^n W_n +\e^m{\bf K}\bstar R_\phi.}\Eq(3.22)$$

The expressions of $A_f$ and $A_\phi$ show that it is convenient to
choose
$K\ge 2m-1$ in order to get them bounded as $\e\to 0$.

The construction of the solution
$R_f$,
$R_\phi$ of
\equ(3.20) is obtained using a fixed point argument to handle the
nonlinear
terms in the equations.

In Appendix A we shall sketch the proof of the Proposition below, which
extends the similar result proved for the one-component Boltzmann gas
without
long-range interactions in [Ca80] and [Lac]. We assume for simplicity
periodic
boundary conditions, namely $\O$ is the $3$-dimensional torus of unit
side.
The potential of the long range force is assumed $C^\infty$, non
negative and of
compact support. Of course such assumptions could be relaxed, but we
will not try
to examine the most general setup. Moreover we use the norm
$$\|f\|_{\a,\ell,s}=\sup_{v\in \IR^{3}}\Big[\hbox{\rm
e}^{\a v^{2}}(1+|v|^2)^{\ell/2}|f(\,\cdot\, , v)|_s \Big]\Eq(3.25)$$
and $|f|_s$ is the Sobolev norm of order $s$.

We will refer below to sufficiently smooth solutions to the VE equations
meaning solutions
which are in $H_s$ for some sufficiently large $s$ and such that the
inequalities
$$T_0\le T\le T_1, \quad \rho_0\le \rho\pm \f\le \rho_1$$
are verified for suitable positive constants $T_0$, $T_1$, $\rho_0$ and
$\rho_1$.
\vskip.3cm
\noindent{\bf Proposition 4.1}. {\it Suppose that $(\r,u,T,\f)$ is a
solution to the Vlasov-Euler equations \equ(3.18), \equ(3.19)
sufficiently smooth in
the time interval $[0,\bar t]$. Then there are positive constants $\e_0$
and $C$
such that, for $\e<\e_0$ a unique classical solution to the system
\equ(3.20)
with $m\ge 4$ exists and satisfies the bounds
$$\eqalign{&\supt\|R_f(\,\cdot\, ,t)\|_{\a,\ell,s}\le C\e
\supt\big[\|A_f(\,\cdot\, ,t)\|_{\a,\ell,s}+\|A_\phi(\,\cdot\,
,t)\|_{\a,\ell,s}\big],\cr&
\supt\|R_\phi(\,\cdot\, ,t)\|_{\a,\ell,s}\le C\e
\supt\big[\|A_f(\,\cdot\, ,t)\|_{\a,\ell,s}+\|A_\phi(\,\cdot\,
,t)\|_{\a,\ell,s}\big],}\Eq(3.26)$$
for any positive $\a<\bar T/2$, $\bar T\defi\sup_{x\in \O, t\in [0,\bar
t]}
T(x,t)$, $\ell>3$, $s\ge 2$.}
\vskip.3cm
\noindent{\bf Corollary 4.2}. {\it Under the assumptions of Proposition
3.1, for
$\e<\e_0$ there is a
smooth solution $(f^\e_t,\phi^\e_t)$ to the rescaled Vlasov-Boltzmann
equations
\equ(2.7) and moreover, denoting
by
$M_t$ the Maxwellian with parameters evolving according to the Euler
equations, it
satisfies:
$$\sup_{0\le t\le \bar
t}[\|f^\e_t-M_t\|_{\a,\ell,s}+\|\phi^\e_t-{\f_t\over
\r_t}M_t\|_{\a,\ell,s}]\le C\e.
$$}
\vskip.8cm\goodbreak
\numsec= 5
\numfor= 1
\head(5, Navier Stokes correction)
\vskip.2cm
The Navier-Stokes corrections to the hydrodynamical equations on the
Euler scale
are usually obtained by means of a suitable resummation of the Hilbert
series
expansion called the {\it Chapman-Enskog expansion}. For our purposes
it is
convenient to
look at
a modified version of the expansion proposed by Caflisch [Ca87].
\vskip.1cm

We use the notation: for $n\ge 0$, $\hat f_n={\cal P} f_n$, $\bar f_n=
{\cal P}^\perp
f_n$,
$\hat\phi_n={\cal K}\phi_n$, $\bar\phi_n={\cal K}^\perp\phi_n$,
$F_n={\bf K}\bstar f_n$, $ W_n={\bf K}\bstar\phi_n$. The terms in the
expansions
are given as follows: we set $M_s:=M(1,u,T)$;
$$ f_0=\r M_s, \quad \phi_0={\f} M_s,\quad \hat f_1=0,\quad \hat\phi_1=0
\Eq(2.12)$$
$${\cal L}\bar f_1={\cal P}^\bot\Big[D_t f_0 +F_0\cdot\nabla_v
f_0 +W_0 \cdot\nabla_v \phi_0 \Big ]
\Eq(2.14.1)$$
$$\Gamma\bar\phi_1=
-J(\phi_0,\bar f_1)+{\cal K}^\bot\Big[ D_t\phi_0-\e{\f\over \r^2}{\cal
P}[D_t \bar
f_1]
+W_0\cdot\nabla_v f_0 + F_0\cdot\nabla_v \phi_0\Big].\Eq(2.16)$$
$${\cal P}\Big[ D_t(f_0+\e \bar f_1) + F_0\cdot\nabla_v
(f_0 +\e\bar f_1)+W_0\cdot\nabla_v (\phi_0 +\e\bar\phi_1) \Big ]=0
\Eq(2.14)$$
$${\cal K}\Big[ D_t(\phi_0+\e \bar \phi_1)+W_0\cdot\nabla_v
(f_0 +\e\bar f_1)+ F_0\cdot\nabla_v (\phi_0
+\e\bar\phi_1)\Big]=0
\Eq(2.15)$$
$$\eqalign{&{\cal L}\bar f_{2}=-2Q( f_{1},f_1)
\cr&+{\cal
P}^\bot\Big[ D_t ( \bar f_1 +\e \hat f_{2})+
\big[F_0\cdot \n_v f_{1}+F_1\cdot \n_v f_{0} +W_0 \cdot \n_v\phi_{1}+W_1
\cdot
\n_v\phi_{0}\big]\Big ]}
\Eq(2.18)$$
$$\eqalign{&\Gamma\bar\phi_{2}=-\varphi J(M_s, f_{2})
-J(\phi_1,f_1)
+{\cal K}^\bot\Big[ D_t\bar \phi_n+{\f\over \r^2}{\cal P}[D_t \bar
f_1]+\e
D_t\hat\phi_{2}\cr&+
\big[W_0\cdot \n_v f_{1}+W_1\cdot \n_v f_{0} +F_0 \cdot \n_v\phi_{1}+F_1
\cdot
\n_v\phi_{0}\big]\Big ]
}
\Eq(2.20)$$
$${\cal P}\Big[ D_t f_{2} + \sum_{\st (h,h'):\ h\ge 0,h'>0\atop
h+h'=2}
\big[F_h\cdot \n_v f_{h'} +W_h \cdot \n_v\phi_{h'}\big]\Big
]=0
\Eq(2.17)$$
$${\cal K}\Big[ D_t \phi_{2}+\sum_{\st (h,h'):\ h\ge 0,h'>0\atop
h+h'=2}
\big[F_h\cdot \n_v \phi_{h'} +W_h \cdot
\n_vf_{h'}\big]\Big]=0
\Eq(2.19)$$
Before giving conditions defining the higher order terms of the
expansion let us comment about previous conditions:
Equations
\equ(2.12)--\equ(2.19) have to be solved in the order they are written:
\equ(2.14.1) and \equ(2.16) are used to determine $\bar f_1$ and then
$\bar \phi_1$ in
terms of the
hydrodynamical parameters $(\r,u,T,\f)$ and their derivatives; as a
consequence
\equ(2.14) and
\equ(2.15) only involve $(\r,u,T,\f)$ and represent the hydrodynamical
equations we
are
looking for. Because of \equ(2.12) $f_1$ and $\phi_1$ are completely
determined. Then
\equ(2.18)
and \equ(2.20) can be solved to find $\bar f_2$ and $\bar\phi_2$,
depending only on
$f_0,f_1,\phi_0, \phi_1$ and on the hydrodynamical parts of $f_2$ and
$\phi_2$.
Finally,
\equ(2.17) and \equ(2.19) are linear equations in the hydrodynamical
part of $f_2$ and
$\phi_2$ which can be used to determine them.

We notice that the term proportional to $\e$ in \equ(2.16) has been
included to avoid
third
order derivatives in the hydrodynamical equations. This is usually done
in the
standard
Chapman-Enskog expansion by {\it expanding the time derivatives in
powers of
$\e$}.

For $n\ge 2$ we set:
$$\eqalign{&{\cal L}\bar f_{n+1}=-\sum_{\st k,j\ge 1 \atop\st k+j=
n+1}2Q( f_{j},f_k)
\cr&+{\cal
P}^\bot\Big[ D_t ( \bar f_n +\e \hat f_{n+1})+ \sum_{\st (h,h'):\
h,h'\ge 0\atop
h+h'=n}
\big[F_h\cdot \n_v f_{h'} +W_h \cdot \n_v\phi_{h'}\big]\Big
]}
\Eq(2.18')$$
$$\eqalign{&\Gamma\bar\phi_{n+1}=-\varphi J(M_s, f_{n+1})
-\sum_{\st k,j\ge 1 \atop\st k+j= n+1}J(\phi_k,f_j)\cr&
+{\cal K}^\bot\Big[ D_t\bar \phi_n+\e D_t\hat\phi_{n+1}+
\sum_{\st (h,h'):\ h,h'\ge 0\atop h+h'=n}
\big[F_h\cdot \n_v \phi_{h'} +W_h\cdot \n_vf_{h'}\big]\Big]
}
\Eq(2.20')$$
$${\cal P}\Big[ D_t f_{n+1} + \sum_{\st (h,h'):\ h\ge 0,h'>0\atop
h+h'=n+1}
\big[F_h\cdot \n_v f_{h'} +W_h \cdot \n_v\phi_{h'}\big]\Big
]=0
\Eq(2.17')$$
$${\cal K}\Big[ D_t \phi_{n+1}+\sum_{\st (h,h'):\ h\ge 0,h'>0\atop
h+h'=n+1}
\big[F_h\cdot \n_v \phi_{h'} +W_h \cdot
\n_vf_{h'}\big]\Big]=0
\Eq(2.19')$$
The procedure used for $f_2$ and $\phi_2$ can be repeated to get $f_n$
and
$\phi_n$ for any $n>2$.
\vskip.2cm

As for the Euler limit discussed in the previous section, instead of
looking for the
convergence of
the expansion we consider its truncation \equ(2.12.5) where the
functions $f_n$ and
$\phi_n$ are
computed according to the procedure just explained, but setting $f_n=0$,
$\phi_n=0$
for $n\ge
K+1$. The remainders
$R_f$ and
$R_\phi$ have to be solutions of the equations
$$\eqalign{D_t R_f + F\cdot\nabla_v R_f + W\cdot\nabla_v
R_\phi=&\e^{-1}{\cal
L}R_f+{\cal L}^{(1)}R_f+\e^{m-1}\big[ J(R_f,R_f)+ A_f \big],
\cr D_t R_\phi+ W\cdot\nabla_v R_f+ F\cdot\nabla_v
R_\phi=&\e^{-1}[\Gamma R_\phi
+\tilde\Theta R_f]+\G^{(1)}R_\phi
\cr&+\e^{m-1}\big[J(R_\phi,R_f)+A_\phi)\big]}\Eq(2.26)$$
where $F$ and $W$ are given by \equ(3.22), ${\cal L}^{(1)}$, $\G^{(1)}$
and
$\tilde\Theta$
are given by \equ(3.21), while the expressions of $A_f$ and $A_\phi$ are
slightly
different
from \equ(3.21.1), and are:
$$\eqalign{&A_f=\e^{K-2m+1}\Big(\sum_{\st (h,h'):\ h, h'\ge 1\atop \st
h+h'>
K+1}\e^{h+h'-K-1}Q(f_h,f_{h'})-\cr&\sum_{\st (h,h'):\ h,h'\ge
0\atop h+h'> K}\e^{h+h'-K}\big[F_h\cdot \n_v f_{h'} +W_h \cdot
\n_v\phi_{h'}\big]\cr&
-D_t \bar f_{K}- {\cal P}[F_0\cdot \n_v f_K+W_0\cdot
\n_v\phi_K]\Big),}\Eq(3.27.1)$$
$$\eqalign{&A_\phi=\e^{K-2m+1}\Big(
\sum_{\st (h,h'):\ h, h'\ge 1
\atop \st h+h'> K+1}\e^{h+h'-K-1}
J(\phi_h,f_{h'})\cr&-\sum_{\st (h,h'):\ h,h'\ge
0\atop h+h'> K}\e^{h+h'-K}
\big[W_h\cdot \n_v f_{h'} +F_h \cdot \n_v\phi_{h'}\big]\cr&-D_t\bar
\phi_{K}-
{\cal K}[W_0\cdot \n_v f_K+F_0\cdot \n_v\phi_K]\Big).}\Eq(3.27.2)
$$
\vskip.3cm
We now exploit the equations \equ(2.14.1)--\equ(2.15) in order to obtain
the
Navier-Stokes
equations for the binary mixture with long range forces.

Note that in \equ(2.14.1) the force terms do not contribute because
${\cal P}^\perp
\n_v M_s=0$.
Therefore $\bar f_1$ has the same expression as for the one component
gas without
self-interaction: recall that
$${\cal P}^\perp(D_tf_0)=M\left[\sum_{i,j=1}^3 A_{i,j}\pt_i u_j
+\sum_{i=1}^3 B_i
\pt_i
T\right],\Eq(5.1.3)$$ with
$$A_{i,j}={1\over T}(\tilde v_i\tilde v_j -{\tilde v^2\over 3}\d_{i,j}),
\quad
B_i=\left({\tilde
v^2\over 2} -{5\over 2}T\right){\tilde v_i\over T^2}$$
and $\tilde v=v-u$.

Therefore
$$\bar f_1=-\psi_1\sum_{i,j=1}^3 A_{i,j}\pt_i u_j -\psi_2\sum_{i=1}^3
B_i \pt_i
T,\Eq(5.1.5)$$
with $\psi_1$ and $\psi_2$ non negative smooth functions of $|\tilde
v|$.
Moreover, ${\cal P}[\n_v \bar f_1]=0$ because $\bar f_1$ is orthogonal
to the
invariants. On
the other hand $(M_s\chi_\a, \n_v \bar\phi_1)=0$ for $\a=0,\dots, 3$,
because
$\bar\phi_1$ is
orthogonal to the constants. Furthermore
$$\int dv [ F_0\cdot\nabla_v (f_0 )+W_0\cdot\nabla_v
\phi_0]=0,\Eq(4.7.5)$$
$$\int dv \tilde v [ F_0\cdot\nabla_v (f_0 )+W_0\cdot\nabla_v
\phi_0=-F_0\rho
-W_0\varphi.\Eq(4.7)$$
Hence the mass equation is
$$\pt_t \r +\n\cdot (\r u)=0,\Eq(4.7.8)$$
and the momentum equation is
$$\rho D^u_t u +\nabla P= -\e\nabla \cdot\sigma +\r F_0 +\f
W_0\Eq(4.8)$$
with $D^u_t=\pt _t+u\cdot\nabla$ and
$$\sigma_{i,j}:=-\nu(\partial_ju_i+\partial_i u_j-{2\over
3}\delta_{i,j}\nabla\cdot
u)$$
and
$$\nu=\int dv \psi_1(|\tilde v|)A_{1,2}^2.$$
In order to compute the equation for the energy and for the
concentration we need the
expression of
$\bar
\phi_1$, which has to be computed using \equ(2.16). This implies
$$\bar\phi_1=\Gamma^{-1}{\cal K}^\bot
\Bigg[ D_t\phi_0 -\e{\f\over \r^2}{\cal P}[D_t \bar f_1]+ W_0\cdot
\nabla_v f_0+ F_0
\cdot
\nabla_v
\phi_0 -J(\phi_0,\bar f_1)\Bigg]. \Eq(5.3)$$
We are interested in computing the component of $\n_v \bar \phi_1$
along $M\chi_4$
i.e., after an integration by parts:
$$(M\chi_4, \n_v \bar \phi_1)=-\int d\tilde v \tilde v \bar
\phi_1.\Eq(5.4).
$$
Since
$J(\phi_0,\bar f_1)={\f\over \r} {\cal L} \bar f_1 -\Gamma \bar f_1$, we
get
$$\Gamma^{-1}J(\phi_0,\bar f_1)={\f\over \r}\Gamma^{-1}{\cal L}\bar f_1
- \bar f_1=
{\f\over \r}\Gamma^{-1}{\cal P}^\perp(D_tf_0) - \bar f_1,$$
after using \equ(2.14.1) to get the second equality
The second term does not contribute to \equ(5.4) since $\bar f_1$ is
orthogonal to $M\tilde v$.
Hence, by \equ(5.1.5),
$$\int dv\tilde v\Gamma^{-1}J(\phi_0,\bar f_1)={\f\over \rho}\int
d\tilde v\tilde
v\G^{-1}\big[\tilde v\cdot \nabla T {M\over 2T^2}(\tilde v^2 - 5
T)\big];\Eq(5.10)$$
the term proportional to $\n u$ is odd in $\tilde v$ and hence does not
contribute to
the previous expression.
Now we compute
$${\cal K}^\bot [D_t(\varphi M_s)+\rho
W\cdot\nabla_vM_s+\phi F\cdot\nabla_vM_s] =$$
$$
\varphi D^u_t(
M_s)+\varphi\tilde v\cdot
\nabla M_s+M_s\tilde v\cdot\nabla\varphi +\rho W\cdot\nabla_vM_s+\f
F\cdot\nabla_vM_s
\Eq(4.3.9)$$
We have
$$\varphi D^u_t(M_s)= {M_s\over 2T^2}\varphi D^u_tT(\tilde v^2
-3T]-\nabla_v M_s
\varphi D^u_t u$$
Since the first term is even in $\tilde v$,
only the second term contributes to $\int dv \tilde
v\Gamma^{-1}_M$ in \equ(5.10).
By using the equation for the momentum we get
$$D^u_t u={1\over \rho}[-\nabla P +\rho F+\varphi W] -{\e\over
\r}\n\cdot
\s\Eq(4.3.11)
$$
Moreover
$$\big[{\f\over \rho}\nabla P-\f F-{\varphi^2\over \rho}W+\rho W+\varphi
F\big]\cdot\nabla_v M_s=
[{\f\over \rho}\nabla P+W{\rho^2-\f ^2\over\rho}]\cdot\nabla_v M_s.$$
Now
$$\eqalign{-{\f\over \rho}\nabla P\cdot{\tilde v\over T}M_s+M_s\tilde
v &\cdot\nabla\varphi=[-\f{1\over T}\nabla T-{\f\over\rho}\nabla
\rho+\nabla\varphi]\cdot
M_s\tilde v=\cr
& [-\f{1\over T}\nabla T+\rho\nabla {\f\over\rho}]\cdot
M_s\tilde v
}$$
and
$$\eqalign{-\f{1\over T}\tilde v\cdot \nabla T
M_s& +\f\tilde v \cdot\nabla M_s=[-{1\over T}\nabla T +({\tilde v^2\over
2T^2}-{3\over 2 T})]\f\tilde v\cdot \nabla T
M_s + {\f\over T}M_s \tilde v \otimes \tilde v\cdot \n u\cr
&=\f\nabla T\cdot
M_s\tilde v [{\tilde v^2\over
2T^2}-{5\over 2 T}]+ {\f\over T}M_s \tilde v \otimes \tilde v\cdot \n u.
}\Eq(10)$$
The last term in \equ(10) does not contribute to \equ(5.4).
Since the first term in the r.h.s of \equ(10) cancels out with the term
in \equ(5.10)
(remember
that $M=\rho M_s$) we have
$$\hbox{\rm r.h.s \equ(4.3.9)}=M\tilde v\Big[\nabla{\f\over\rho}-
{W\over\rho^2
T}(\rho^2-\f^2)
-\e{\f\over\rho^2}\n\cdot \s\Big].\Eq(11)$$
It is now easy to check that in the computation of the l.h.s. of
\equ(5.3) the last
term of
\equ(11) is compensated by the term $\e{\f\over
\r^2}{\cal P}[D_t \bar f_1]$. Collecting terms we get
$$(M_s\chi_4,\n_v\bar\phi_1)=-D\Big[\nabla{\phi\over\rho}- {W\over
\rho^2T}(\rho^2-\phi^2)\Big]
\Eq(4.2)$$
where
$$D:=\int dv M\tilde v_i\Gamma^{-1}\tilde v_i
\Eq(4.3)$$
\vskip.5cm
The other terms appearing in the equation for the energy are computed
in the standard
way:
$$\int d\tilde v\chi_4\Bigg[\v\cdot \nabla \bar f_1 +F_0\cdot
\nabla_v\bar f_1
+W_0\cdot\nabla_v
\bar
\phi_1\Bigg]=0
\Eq(4.5.1)$$
$$\nabla \cdot\int d\tilde v\chi_4 \bar f_1 =-\n\cdot [\kappa \n T],$$
with
$$\kappa=\int d\tilde v \psi_2B_i^2.$$
Moreover
$$\int dv\chi_4F_0\cdot\nabla_v\bar f_1=-\int dv \tilde v\cdot F_0\bar
f_1=0$$
$$\int dv\chi_4W_0\cdot\nabla_v\bar\phi_1=-\int dv \tilde v\cdot
W_0\bar\phi_1=
-W_0\cdot DQ,$$
with
$$Q:=\nabla{\f\over\rho} - {1\over \rho^2T} W (\rho^2-\f ^2)\Eq(4.5)$$
Therefore, the equation for the energy is

$${3\over 2} \rho D_t T +p \nabla\cdot u=\e\nabla(\kappa\nabla T)
-\e\sigma: \nabla u
-\e W\cdot
D Q
\Eq(4.6)$$

\vskip.2cm
Finally, to get the equation for the concentration we have to
exploit the condition \equ(2.15).

$$\eqalign{
{\cal K}&\Big[ D_t(\varphi M+\e \bar \phi_1)\Big]=\partial_t\varphi+\int
dv
v\cdot\nabla
\varphi M +\e \int dv v\cdot\nabla \bar\phi_1
\cr= & \partial_t\varphi+ \nabla\cdot(u\f)+\e\nabla\cdot\int dv\tilde
v\bar\phi_1 }
\Eq(4.1.1)
$$
The last term has already been computed in \equ(4.2). Therefore, the
equation for
$\f$ is
$$\partial_t \varphi + \nabla\cdot(\varphi u) =\e \nabla\cdot(D
Q)\Eq(4.4)$$
where $Q$ is given in \equ(4.5).

Recalling the definitions of $F_0$ and $W_0$
we finally get the {\it Vlasov-Navier-Stokes equations} (VNS) for a
binary mixture given by \equ(4.11).
\vskip.3cm
As for the Euler limit, the arguments in the Appendix A prove the
following
Proposition 5.1
which holds under the same assumptions as before Proposition 3.1:
periodic boundary
conditions and
smoothness of the long range potential.
\vskip.3cm
\noindent{\bf Proposition 5.1}. {\it Suppose that for $\e>0$ small
enough
there is a solution $(\r^\e,u^\e,T^\e,\f^\e)$ to the
Vlasov-Navier-Stokes equations
\equ(4.11) sufficiently smooth in the time interval $[0,\bar t]$
independent of $\e$.
Then there are
positive constants $\e_0$ and $C$ such that, for $\e<\e_0$ an unique
classical
solution to the system
\equ(2.26) with $m\ge 4$ exists and satisfies the bounds
$$\eqalign{&\supt\|R_f(\,\cdot\, ,t)\|_{\a,\ell,s}\le C\e
\supt\big[\|A_f(\,\cdot\, ,t)\|_{\a,\ell,s}+\|A_\phi(\,\cdot\,
,t)\|_{\a,\ell,s}\big],\cr&
\supt\|R_\phi(\,\cdot\, ,t)\|_{\a,\ell,s}\le C\e
\supt\big[\|A_f(\,\cdot\, ,t)\|_{\a,\ell,s}+\|A_\phi(\,\cdot\,
,t)\|_{\a,\ell,s}\big],}\Eq(5.26)$$
for any $\a<\bar T/2$, $\bar T\defi\sup_{x\in \O, t\in [0,\bar t]}
T^\e(x,t)$, $\ell>3$, $s\ge 2$.}
\vskip.3cm
\noindent{\bf Corollary 5.2}. {\it Under the assumptions of Proposition
3.1, for
$\e<\e_0$ there is a
smooth solution $(f^\e_t,\phi^\e_t)$ to the rescaled Vlasov-Boltzmann
equation
\equ(2.7) and moreover, denoting
by
$M_t$ the Maxwellian with parameters evolving according to the
Vlasov-Navier-Stokes
equations, it satisfies:
$$\sup_{0\le t\le \bar t}[\|f^\e_t-M_t-\e
f_1\|_{\a,\ell,s}+\|\phi^\e_t-{\f_t\over
\r_t}M_t-\e\phi_1\|_{\a,\ell,s}\le C\e^2].
$$}
\vskip.8cm
\numsec=6
\numfor=1
\head(6, Incompressible Navier-Stokes limit)
\vskip.5cm
In this Section we consider a different scaling limit such that one
can get hydrodynamic equations with
dissipative terms of order $1$ instead of order $\e$ as in the
previous Section. We choose also in this case $\gamma=\e$ but
we use the parabolic space time scaling

$$t\to \e^{-2}t,\quad x\to \e^{-1}x. $$

After rescaling,
eq.'s \equ(2.5') become:
$$\pt_t f +\e^{-1}v\cdot \n_x f +\e^{-1}F\cdot \n_v f+ \e^{-1}W\cdot
\n_v\phi =
\e^{-2}J(f,f),\Eq(6.0)$$
$$\pt_t \phi +\e^{-1}v\cdot \n_x \phi +\e^{-1}F\cdot \n_v \phi+\e^{-1}
W\cdot
\n_vf = \e^{-2}J(\phi,f),$$
with $F$ and $W$ given by \equ(2.8).

We shall solve \equ(6.0) as in the Euler case in terms of a truncated
Hilbert
expansion of the form
$$\eqalign{&f=\sum_{n=0}^K\e^n f_n+\e^m R_f,\cr&\phi=\sum_{n=0}^{K}\e^n
\Phi_n+\e^m R_\Phi,}\Eq(6.1.1)$$
with suitably chosen positive integers $K$ and $m$, but in this case
we choose in a different way the terms of order $0$ in the expansion.

$$f_0= M_0, \quad \phi_0= 0$$
where $ M_0$ is a Maxwellian with parameters $\bar\rho$ and $\bar T$
some fixed
constants and $u=0$. This implies the
vanishing of the forces at the lowest order: $W_0=F_0=0$. We remark that
choosing the first order term in the expansion to be a global
Maxwellian is essential to the incompressible limit set up. The choice
$\phi_0=0$ is made for simplifying the computations.
By plugging \equ(6.1.1) in the rescaled Boltzmann equations \equ(6.0),
one easily obtains
the conditions for the higher order terms in the
expansions:
$${\cal L}f_1=0,
\Eq(6.2)$$
$$\Gamma \phi_1=0,\Eq(6.2.1)$$
for $0\le n\le K-1$:
$$\eqalign{&{\cal L}f_{n+2}+\sum_{\st (h,h'):\ h, h'\ge 1
\atop \st h+h'=n+2}Q(f_h,f_{h'})=\cr&\sum_{\st (h,h'):\ h>0,h'\ge 0\atop
h+h'=n+1}
\big[F_h\cdot \n_v f_{h'} +W_h \cdot \n_v\phi_{h'}\big]+v\cdot \n_x
f_{n+1}+\pt_t f_{n},}\Eq(6.2.2)$$
$$\eqalign{&\G\phi_{n+2}+ \sum_{\st (h,h'):\ h, h'\ge 1
\atop \st h+h'=n+2}J(\phi_h,f_{h'})\cr&=\sum_{\st (h,h'):\ h,h'\ge
0\atop h+h'=n+1}
\big[F_h\cdot \n_v \phi_{h'} +W_h \cdot \n_vf_{h'}\big]+v\cdot \n_x
\phi_{n+1}+\pt_t
\phi_{n}.}\Eq(6.2.3)$$
Moreover
$$\eqalign{&\pt_t R_f +\e^{-1}\left[v\cdot R_f+ F\cdot\n_v R_f+W\cdot
\n_v
R_\phi\right] \cr&=\e^{-2} {\cal L}R_f+\e^{-1} {\cal L}^{(1)}R_f+{\cal
L}^{(2)}R_f+\e^{m-2}[J(R_f,R_f)+A_f],
\cr& \pt_t R_\phi +\e^{-1}\left[v\cdot R_\phi +F\cdot\n_v R_\phi+W\cdot
\n_v
R_f\right]=\e^{-2} \G R_\phi\cr&
+\e^{-1}\G^{(1)}R_\phi+\e^{-1}\tilde\Theta^{(1)}
R_f+\G^{(2)}R_\phi+\tilde\Theta^{(2)} R_f+
\e^{m-2}[J(R_\phi,R_f)+A_\phi],}\Eq(6.20)$$
where
$$\eqalign{&{\cal L}^{(1)} g= J(f_1,g)+J(g,f_1),\quad{\cal L}^{(2)} g =
\sum_{h=2}^K
\e^{h-2}[J(f_h,g)+J(g,f_h)],\cr&
\tilde\Theta^{(1)} g=J(\phi_1,g),\quad \tilde\Theta^{(2)} g=
J(\sum_{n=2}^K\e^{n-2}
\phi_n, g),\cr&
\G^{(1)}g=J(g,f_1),\quad \G^{(2)} g = \sum_{h=2}^K
\e^{h-2}J(g,f_h),}\Eq(6.21)$$
$$\eqalign{&A_f=\e^{K-2m+2}\Big(
\sum_{\st (h,h'):\ h, h'\ge 1
\atop \st h+h'> K+2}\e^{h+h'-K-2}
Q(f_h,f_{h'})\cr&-\sum_{\st (h,h'):\ h,h'\ge
0\atop h+h'> K+1}\e^{h+h'-K-1}
\big[F_h\cdot \n_v f_{h'} +W_h \cdot \n_v\phi_{h'}\big]\cr&
-\pt_t f_{K-1}-v\cdot \n_x f_K-\e\pt f_K\Big)\cr&
A_\phi=\e^{K-2m+1}\Big(
\sum_{\st (h,h'):\ h, h'\ge 1
\atop \st h+h'> K+2}\e^{h+h'-K-2}
J(\phi_h,f_{h'})\cr&-\sum_{\st (h,h'):\ h,h'\ge
0\atop h+h'> K+1}\e^{h+h'-K-1}
\big[W_h\cdot \n_v f_{h'} +F_h \cdot \n_v\phi_{h'}\big]\cr&
-\pt_t \phi_{K-1}-v\cdot \n_x \phi_K-\e\pt \phi_K\Big),}\Eq(6.21.1)
$$
and $F$ and $W$ are given by
$$\eqalign{&F=\sum_{n=1}^K\e^n F_n+ \e^m{\bf K}\bstar R_f,
\cr& W= \sum_{n=1}^K \e^n W_n +\e^m{\bf K}\bstar R_\phi,}\Eq(6.22)$$
with $F_n={\bf K}\bstar f_n$, $W_n={\bf K}\bstar\phi_n$.
\vskip.3cm
We now find the expressions of the first terms
in the expansions $f_1,\phi_1,f_2,\phi_2$.

\noindent From \equ(6.2.1) we get
$$\phi_1=\f_1 M_0$$
and from \equ(6.2) we have
$$f_1=\ M_0(v)\left(\rho_1+ u_1\cdot {v \over \bar T}+\theta_1{v^2-3\bar
T\over
\bar T^2}\right)$$ where $\rho_1, u_1,\theta_1$ are to be determined as
functions of
$x,t$. By \equ(6.2.2) with $n=0$ we obtain
$${\cal P} ( v\cdot \n_x f_1 +F_1\cdot \n_v f_0
)=0.$$
But
$${\cal P}[v\cdot\n_x f_1]= M_0\left[\left(1+{v^2-3\bar T\over 2 \bar
T^2}\right)\n_x\cdot u_1 +{v\over \bar T}\cdot
\n_x(\bar T\rho_1+\bar\rho\theta_1)\right],$$
while
$${\cal P}[F_1\cdot \n_vf_0]= - M_0{v\over \bar T}\cdot F_1.$$
Hence we find the conditions
$$\n_x\cdot u_1=0,\quad \n_x\Big[\rho_1+\theta_1+\int
d\/y U(|x-y|)\rho_1(y)\Big]=0.
\Eq(6.7)$$
On the other hand,
$${\cal P}^\perp[F_1\cdot \n_v f_0 ]=0,$$
so
$$f_2={\cal L}^{-1} \Big[{\cal P}^\perp v\cdot\n_x f_1\Big] -{\cal
L}^{-1}J(f_1,f_1)
+\hat f_2,$$
with $\hat f_2\in \hbox{\rm Null}{\cal L}$.

Therefore $f_2$ has the usual expression
$$f_2={1\over 2} \sum_{i,j=1}^3 A_{i,j}[ u_{1,i}u_{1,j}-\psi_1\pt_i
u_{1,j}]+
\sum_{i=1}^3 B_i[\theta_1 u_{1,i}-\psi_2 \pt_i\theta_1]+ {1\over 2} M_0
\theta_1^2{\cal P}^\perp\Big[\Big({v^2-3\over 2}\Big)^2\Big].\Eq(6.99)$$

>From \equ(6.2.3) with $n=0$ we get the expression for $\phi_2$:
$$\phi_2= \Gamma^{-1}\Big[ v\cdot \n_x \phi_1+W_0\n_v
f_0-J(\phi_1,f_1)\Big]+\hat \phi_2,$$
with $\hat \phi_2 \in \hbox{\rm Null}{\G}$.

Moreover, by \equ(6.2)
$$\Gamma^{-1}J(\phi_1,f_1)=-\f_1\Gamma^{-1}\Gamma f_1= -\f_1 f_1.$$
Hence
$$\phi_2= \f_1 f_1+ \Big[\n_x \f_1-{1\over \bar T} W_1\Big]\cdot
\Gamma^{-1}[v M_0 ]+\hat \phi_2.$$
>From \equ(6.2.2) with $n=1$ we get

$$\pt_t u_1+u_1\cdot \n_x u_1= -\n_x p + F_2+ \nu\Delta_xu_1+\rho_1 F_1
+\f_1
W_1$$
and
$${5\over 2}[\pt_t \theta_1 +u_1\cdot\nabla_x\theta_1]=u_1\cdot
F_1+{\kappa}\Delta_x\theta_1$$
Since $F_2=\n_x G$, with $G(x)=\int d\/y U(|x-y|)\rho_2(y)$,
putting $\bar p=p-G$, the previous equation reduces to
$$\pt_t u_1+u_1\cdot \n_x u_1= -\n_x \bar p+ \nu\Delta_xu_1+\rho_1 F_1
+\f_1
W_1,$$
which is the usual incompressible Navier-Stokes equation with the
self-consistent force
$$\rho_1 F_1 +\f_1 W_1=-\rho_1\n_x\int d\/y U(|x-y|)\rho_1(y)
+\f_1\n_x\int
d\/y U(|x-y|)\f_1(y).$$
Finally from \equ(6.2.1) with $n=1$ we get the equation for $\f_1$
$$\pt_t\f_1 +u_1\cdot \nabla_x \f_1= D\Big[{1\over\bar\rho}\Delta_x
\f_1-
{1\over\bar T}\Delta_x\int d\/y
U(|x-y|)\f_1(y)\big]\Big],$$
with
$$D= - \int d\/v v\cdot \Gamma^{-1}(vM).\Eq(9)$$

Summarizing, dropping the index $1$, the set of equations for
$\rho, u, \theta,\phi, p$
is:
$$\eqalign{&\pt_t u+u\cdot \n_x u= -\n_x p + \nu\Delta_xu+\rho F +\f
W,\cr&
{5\over 2}[\pt_t \theta +u\cdot\nabla_x\theta]=u\cdot
F+k\Delta_x\theta,\cr&
\pt_t\f +u\cdot\nabla_x \f= D\Big[{1\over\bar\rho}\Delta_x
\f- {1\over\bar T }\Delta_x\int d\/y
U(|x-y|)\f(y)\big]\Big],
\cr&F=-\nabla_x\int dyU(|x-y|)\rho(y),\quad W= \nabla_x\int
dyU(|x-y|)\f(y),
\cr& \n_x\Big[\rho+\theta+\int d\/y
U(|x-y|)\rho(y)\Big]=0,\cr&\nabla_x\cdot
u=0.}\Eq(INS)
$$
The equation for $\f$ is linear unlike the one we get in the VNS
equations, but there is still a non linear term in $\f$ in the
momentum equation. The equation for $\theta$ which corresponds to the
deviation in the
temperature decouples from the rest. In fact,
if we consider a solution to the previous equation with an initial datum
$\rho=const, \theta=const$ such conditions persist in time and $u$ and
$\f$ have
to solve the simplified set of equations
$$\eqalign{&\pt_t u+u\cdot \n_x u= -\n_x p + \nu\Delta_xu +\f W,
\cr& \pt_t\f +u\cdot\nabla_x \f= D\Big[{1\over\bar\rho}\Delta_x
\f- {1\over\bar T }\Delta_x\int d\/yU(|x-y|)\f(y)\big]\Big],
\cr&W=\nabla_x\int dyU(|x-y|)\f(y),
\cr&
\nabla_x\cdot u=0.}\Eq(10a)$$

\vskip.8cm
In Appendix B there is the proof of the following proposition:

\noindent{\bf Proposition 6.1}. {\it Suppose that for $\e>0$ small
enough
there is a solution $(\r,u,T,\f)$ to the incompressible Navier-Stokes
equations
\equ(INS) sufficiently smooth in the time interval $[0,\bar t]$
independent of $\e$.
Then there are
positive constants $\e_0$ and $C$ such that, for $\e<\e_0$ a unique
classical
solution to the system
\equ(6.20) exists and satisfies the bounds
$$\eqalign{&\supt\|R_f(\,\cdot\, ,t)\|_{\a,\ell,s}\le C\e
\supt\big[\|A_f(\,\cdot\, ,t)\|_{\a,\ell,s}+\|A_\phi(\,\cdot\,
,t)\|_{\a,\ell,s}\big],\cr&
\supt\|R_\phi(\,\cdot\, ,t)\|_{\a,\ell,s}\le C\e
\supt\big[\|A_f(\,\cdot\, ,t)\|_{\a,\ell,s}+\|A_\phi(\,\cdot\,
,t)\|_{\a,\ell,s}\big],}\Eq(.26)$$
for any $\a<\bar T/2$, $\ell>3$, $s <m$.}
\vskip.3cm
\noindent{\bf Corollary 6.2}. {\it Under the assumptions of Proposition
3.1, there is
for $\e<\e_0$ a
smooth solution $(f^\e_t,\phi^\e_t)$ to the rescaled Vlasov-Boltzmann
equation
\equ(6.0) and moreover, denoting
by
$M_t$ the Maxwellian with parameters evolving according to the
incompressible
Navier-Stokes equations
\equ(INS), it satisfies:
$$\sup_{0\le t\le \bar t}||f^\e_t-M_0-\e
f_1||_{\a,\ell,s}+||\phi^\e_t-M_0-\e\phi_1||_{\a,\ell,s}\le
C\e^2]
$$}.
\vskip.8cm
{\bf Appendix A}
\numfor=1
\vskip.2cm
We present a sketch of the proof of Proposition 3.1 in the case of hard
spheres, where
$\nu(v)\approx |v|$ for large $v$'s. The extension to more general cross
sections is possible
along the lines proposed in [DE], but we do not discuss it. The
smoothness and decay
properties of the expansion terms are obtained by now standard methods
([Ca80], [DEL],
[ELM94], [ELM95], [ELM98], [ELM99]) which allow to prove the following
\vskip.2cm

{\bf Theorem A.1}

{\it Given $\e>0$, assume that there exists a sufficiently smooth
solution of the Vlasov-Euler equations
\equ(3.18) and \equ(3.19) in the time interval $[0,\bar t]$. Then for
any $j>3$, $s<s_0$ and
$0<\a<T^*=\sup_{(x,t)\in\Omega\times [0,\bar t]}T(x,t)$ there is a
constant $c>0$ such that
the terms in the expansion $f_i,\phi_i,
\ i=1,...,K$, with $K=2m$, solutions of the equations
\equ(2.17)--\equ(2.20)

$$||f_i||_{\a,j,s}\le c,\quad ||\phi_i||_{\a,j,s}\le c
\Eqa(A.1)$$
$$||\partial_vf_i||_{\a,j,2}\le c,\quad ||\partial_v\phi_i||_{\a,j,2}\le
c\Eqa(A.1.1)$$
}
\vskip.2cm
We need bounds on the remainders $R_f$ and $R_\phi$.

We set
$$R^r\equiv R^{(1)}= R_f-R_\phi, \quad R^b\equiv R^{(2)}=R_f+R_\phi.
\Eqa(RR)$$

The reason is that terms like $W\cdot \nabla_v R_\phi$ and $F\cdot
\nabla_v
R_f$ in \equ(3.20) are not well suited when some force term is present
to represent
the
solutions of the equations in terms of characteristics, which is
essential in the
method we are
going to use. The equations for the new variables are:
$$\eqalign{
D_t R^{(1)}& + F^{(1)}\cdot\n_v R^{(1)}=\e^{-1} \Big[\rho^{(1)}\bar{\cal
L}R^{(1)}+\rho^{(1)}\bar
\Theta R^{(2)} +\rho^{(2)}\bar \Gamma R^{(1)}\Big]
+{\cal L}_1 R^{(1)} \cr &
+ \Gamma_2 R^{(1)} + +\Theta_1
R^{(2)} +
\e^{m-1}\big[J(R^{(1)},R^{(1)})+ J(R^{(1)},R^{(2)})+A^{(1)}\big],\cr
D_t R^{(2)}& + F^{(2)}\cdot\n_v R^{(2)}=
\e^{-1} \Big[ \rho^{(2)}\bar{\cal L}R^{(2)}+\rho^{(2)}\bar \Theta
R^{(1)} +\rho^{(1)}\bar \Gamma R^{(2)}\Big] +
{\cal
L}_2 R^{(2)}\cr &
+ \Gamma_1 R^{(2)} +\Theta_2 R^{(1)}+
\e^{m-1}\big[J(R^{(2)},R^{(2)})+ J(R^{(2)},R^{(1)})+A^{(2)}\big],
}\Eqa(A.1.2)$$
where $\bar M$ is the Maxwellian $M$ with $\rho=1$ and
$$\eqalign{
& f_j= {1\over 2}[f^{(1)}_j+ f^{(2)}_j],\quad \phi_j= {1\over
2}[-f^{(1)}_j+
f^{(2)}_j]\cr &
\bar{\cal L} h=J( \bar M,
h)+J(h,\bar M),\cr
&\bar \Theta h=J( \bar M,h),\quad \bar\Gamma h= J(h,\bar M)\cr
&
{\cal L}_ih=J(\sum_{j=1}^K
\e^{j-1}f^{(i)}_j,h)+J(h,\sum_{j=1}^K\e^{j-1}
f^{(i)}_j)\cr
&\Gamma_i h=J(h,\sum_{j=1}^K\e^{j-1} f^{(i)}_j), \quad \Theta_ih
=J(\sum_{j=1}^K
\e^{j-1}f^{(i)}_j,h)\cr
&
A^{(1)}=A_f+A_\phi, \quad A^{(2)}=A_f-A_\phi\cr
}
$$

Following Caflisch [Ca80] we now decompose the remainders in low
and high velocity parts, by looking
for solutions to equations \equ(A.1.2) in the form
$$R^{(1)}=\sqrt{\rho^{(1)}\bar M} g^{(1)} +\sqrt{M^*} h^{(1)},\quad
R^{(2)}=\sqrt {\rho^{(2)}\bar M} g^{(2)}
+\sqrt {M^*} h^{(2)}$$
$M^*$ is a global Maxwellian with a
temperature
$T^*$.
We have
$$\eqalign{
D_t g^{(1)}& +F^{(1)}\cdot\nabla g^{(1)} =\e^{-1} \Big[\rho^{(1)}\bar L
g^{(1)}+\sqrt{\rho^{(1)}}\sqrt{\rho^{(2)}}\bar
T g^{(2)}+\rho^{(2)}\bar G g^{(1)} \Big] \cr
& \hskip2cm +
{\e}^{-1} \chi{\s} ^{-1} \Big[\sqrt{\rho^{(1)}}(K^* h^{(1)}
+ K^*_T h^{(2)})+ {\rho^{(2)}\over\sqrt{\rho^{(1)}}}K^*_G h^{(1)}\Big]
\cr D_t h^{(1)}& +F^{(1)}\cdot\nabla h^{(1)}=\sigma[\mu^{(1)} + F^{(1)}
\cdot
\mu'^{(1)} ] \sqrt{\rho^{(1)}}g^{(1)}+F^{(1)}
\cdot \mu'_* h^{(1)}\cr
&+
{\e}^{-1}\rho^{(1)}\Big[-\nu + \bar\chi (K^* h^{(1)}+ K^*_T h^{(2)})
+{\rho^{(2)}\over\rho^{(1)}}(-\nu_G+K^*_G h^{(1)})\Big]
\cr
&
+L_1\big(\s \sqrt{\rho^{(1)}}g^{(1)}+h^{(1)}\big)+G_2\big(\s
\sqrt{\rho^{(1)}}g^{(1)}+h^{(1)}\big)+T_1\big(\s\sqrt{\rho^{(2)}}
g^{(2)}+h^{(2)}\big)\cr &
+\e^{m-1}\Big[\nu^*Q^*\big(\s \rho^{(1)}g^{(1)}+h^{(1)},\s\rho^{(1)}
g^{(1)}+h^{(1)}\big)\cr &+
\nu^*Q^*\big(\s \rho^{(1)}g^{(1)}+h^{(1)},\s
\rho^{(2)}g^{(2)}+h^{(2)}\big)+A^{(1)}\Big] }
\Eqa(A.4)$$

The equation for $g^{(2)},h^{(2)}$ is obtained by the exchange $1\to 2$.

where
$$\chi(v)=\cases{1,\quad |v|\le \gamma\cr
\hskip3cm\bar\chi=1-\chi\cr
0,\quad \hbox{ otherwise}}
\Eqa(A.5)$$

$$\eqalign{
\mu^{(i)}&={1\over 2} D_t (\log \bar M),\quad \mu'^{(i)}={1\over
2}\rho^{(i)}\nabla_v
\log\bar M, i=1,2\cr&
\mu'_*={1\over 2}\nabla_v \log M_*,\quad
\s=\sqrt{\bar M\over M^*}
}
\Eqa(A.6)$$

$$\eqalign{
&\bar L f={1\over\sqrt {\bar M}}\bar {\cal L}\sqrt {\bar M} f)=(-\nu +K)
f, \quad
L^* f={1\over\sqrt {M^*}}{\cal L}\sqrt {M^*} f)=-\nu^* +K^* f\cr
&\bar T f={1\over\sqrt {\bar M}}\bar \Theta \sqrt {\bar M} f,\quad \bar
G
f={1\over\sqrt {\bar
M}}\bar \Gamma\sqrt {\bar M}f=(-\nu_G +\bar K_G) f,,
\quad K^*_T={1\over\sqrt {M^*}}\bar \Theta \sqrt {M^*} f\cr
&G^* f={1\over\sqrt M^*}\bar \Gamma\sqrt {M^*} f=(-\nu^*_G +K^*_G)f,
\quad
G_i f={1\over\sqrt {M^*}}\Gamma_i\sqrt {M^*} f, i=1,2\cr
& L_i f={1\over\sqrt {M^*}}{\cal L}_i\sqrt {M^*} f, i=1,2, \quad
T_i f={1\over\sqrt {M^*}}\Theta_i\sqrt {M^*} f, i=1,2\cr
}
\Eqa(A.7)$$

$$
\nu^* Q^*(f,\ell)={1\over\sqrt {M^*}}Q(\sqrt {M^*}f,\sqrt
{M^*}\ell),\quad
\nu^* J^*(f,\ell)={1\over\sqrt {M^*}}J(\sqrt {M^*}f,\sqrt {M^*}\ell)\
\Eqa(A.8)$$
\vskip.7cm
{\it Linear problem}
\vskip.2cm
We solve first the linear problem for $g_i, h_i, i=1,2$, assuming that
$F^{(i)}$ are given
functions such that
$$\|F^{(i)}\|_\infty+\|\n_x F^{(i)}\|_\infty<\a_F.\Eqa(bound)$$
We consider the linear system
$$\eqalign{
D_t g^{(1)}& +F^{(1)}\cdot\nabla g^{(1)} =\e^{-1} \Big[\rho^{(1)}\bar L
g^{(1)}+\sqrt{\rho^{(1)}}\sqrt{\rho^{(2)}}\bar
T g^{(2)}+\rho^{(2)}\bar G g^{(1)} \Big] \cr
& +
{\e}^{-1} \chi{\s} ^{-1} \Big[\sqrt{\rho^{(1)}}(K^* h^{(1)}
+ K^*_T h^{(2)})+ {\rho^{(2)}\over\sqrt{\rho^{(1)}}}K^*_G h^{(1)}\Big]
\cr D_t h^{(1)}& +F^{(1)}\cdot\nabla h^{(1)}=\sigma[\mu^{(1)} + F^{(1)}
\cdot
\mu'^{(1)} ] \sqrt{\rho^{(1)}}g^{(1)}+F^{(1)}
\cdot \mu'_* h^{(1)}\cr
&+
{\e}^{-1}\rho^{(1)}\Big[-\nu + \bar\chi (K^* h^{(1)}+ K^*_T h^{(2)})
+{\rho^{(2)}\over\rho^{(1)}}(-\nu_G+K^*_G h^{(1)})\Big]
\cr
&
+L_1\big(\s \sqrt{\rho^{(1)}}g^{(1)}+h^{(1)}\big)+G_2\big(\s
\sqrt{\rho^{(1)}}g^{(1)}+h^{(1)}\big)+T_1\big(\s\sqrt{\rho^{(2)}}
g^{(2)}+h^{(2)}\big)\cr &
+\e^{m-1}D^{(1)} }
\Eqa(A.9)$$
and the equation for $g^{(2)},h^{(2)}$ obtained by the exchange
$1\to 2$. Here $ F^{(1)}$ has to be considered as a given force.

We use the integral form of \equ(A.9) [ELM98]:
$$\eqalign{
g^{(i)}(t,x,v)&= \int_{t^-}^t ds
H^{(i)}\big(s,\f^{(i)}_{s-t}(x,v)\big)\exp\Big\{-\int_{s}^t ds'{1\over
\e}
\tilde\nu^{(i)}\big(\f^{(i)}_{s'-t}(x,v)\big)\Big\}. }\Eqa(A.10)
$$
where $\tilde\nu^{(i)}= \r^{(i)}\nu+ \r^{(j)}\nu_G$, $\f_t^{(i)}(x,v)$
the characteristics of
the equation
$$\pt_t f +v\cdot \nabla_x f + F^{(i)}\cdot\nabla_{v}f=0
\Eqa(A.11)$$
and
$$\eqalign{H^{(1)}=&\e^{-1} \Big[\rho^{(1)}K
g^{(1)}+\sqrt{\rho^{(1)}}\sqrt{\rho^{(2)}}\bar
T g^{(2)}+\rho^{(2)}K_G g^{(1)} \Big] +\cr&
{\e}^{-1} \chi{\s} ^{-1} \Big[\sqrt{\rho^{(1)}}(K^* h^{(1)}
+ K^*_T h^{(2)})+ {\rho^{(2)}\over\sqrt{\rho^{(1)}}}K^*_G h^{(1)}\Big]}
\Eqa(A.12)$$
and $H^{(2)}$ is given by the same expression after the exchange $1\to
2$.
$$\eqalign{
h^{(i)}(t,x,v)&= \int_{t^-}^t ds
H'^{(i)}\big(s,\f{(i)}_{s-t}(x,v)\big)\exp\Big\{-\int_{s}^t ds'{1\over
\e}
\hat\nu^{(i)}\big(\f{(i)}_{s'-t}(x,v)\big)\Big\}. }
\Eqa(A.13) $$
with
$$\hat\nu^{(i)}= \tilde \nu^{(i)}- \e \mu'_*\cdot F^{(i)}$$
(which is positive for $\e$ sufficiently small, depending on $\a_F$,
since $\tilde \nu^{(i)}$
grow linearly at high velocities) and
$$\eqalign{
H'^{(1)}&=
{\e}^{-1}\rho^{(1)}\Big[ \bar\chi (K_* h^{(1)}+ K^*_T h^{(2)})
+{\rho^{(2)}\over\rho^{(1)}}(K^*_G h^{(1)})\Big]\cr
&
+{ L}^{(1)}(\s\sqrt{\rho^{(1)}}
g^{(1)}+h^{(1)})+G^{(2)}(\s\sqrt{\rho^{(1)}}
g^{(1)}+h^{(1)})+T^{(1)}(\s \sqrt{\rho^{(2)}}
g^{(2)}+h^{(2)})\cr
&
+\sigma[\mu^{(1)} + F^{(1)} \mu'^{(1)} ]
\sqrt{\rho^{(1)}}g^{(1)}+\e^{m-1}D^{(1)} }
\Eqa(A.14)$$

We do not write explicitly the equations for $g^{(2)},h^{(2)}$ in
integral form. In
the
following we use the compact notation: $g=\{g^{(1)},g^{(2)}\}$ and
$h=\{h^{(1)},h^{(2)}\}$.
Below we use the notation
$\|\,\cdot\,\|_{\ell,s}=\|\,\cdot\,\|_{0.\ell,s}$ and
$\|\,\cdot\,\|_{\ell}=\|\,\cdot\,\|_{0.\ell,0}$.
Generalizing the method
by Caflisch [Ca80]to our case, we get bounds for the norms
$\|\,\cdot\,\|_{\ell}$ of
$g^{(i)}$ $,h^{(i)}$ in terms of the
$L_2$ norm of $g^{(i)}$ in the form
$$\eqalign{
||h||_r&\le\e(1+\a_F) ||g||_{L_2}+\e^m|D|_{r-1}\cr
||g||_r&\le ||g||_{L_2}+\e^{m+1}|D|_{r},
}
\Eqa(A.16)
$$
provided that $\e,\e_0$ for some suitable $\e_0$ positive and finite for
any finite $\a_F$.
To conclude the argument we need a bound for
$||g||_{L_2}=\sum_{i=1}^2||g^{(i)}||_{L_2}$ in terms of the $L_2$ norm
of $D$.
This last
step is not standard so that we give a sketch of the proof.

To estimate $||g||_{L_2}$, we multiply the first equation in \equ(A.9)
by
$g^{(i)}$,
$i,j=1,2, i\ne j$, respectively, integrate over
$x,v$ and finally sum over $i=1,2$
$$\eqalign{
{1\over 2}
{d\over dt}
&
\Big[||g^{(1)}||_{L_2}^2+||g^{(2)}||_{L_2}^2\Big]
=\e^{-1}
\Big[\big<\sqrt{\rho^{(1)}}g^{(1)}, L \sqrt{\rho^{(1)}}g^{(1)}\big>+
\big<\sqrt{\rho^{(2)}}g^{(2)},
L \sqrt{\rho^{(2)}}g_2\big> \Big]\cr
&
+\e^{-1}
\Big[
\Big(
\big<\sqrt{\rho^{(2)}}g^{(1)},
{\bar M}^{-1/2}J(\bar M,\sqrt{\bar M}
\sqrt{\rho^{(1)}}g^{(2)})\big> +\cr&
\big<\sqrt{\rho^{(1)}}g^{(2)},{{\bar M}}^{-1/2}J(\bar
M,\sqrt{\bar M}
\sqrt{\rho^{(2)}}g^{(1)}\big>\Big)+
\cr&
+\big<\sqrt{\rho^{(2)}}g^{(1)},{{\bar M}}^{-1/2}J(\sqrt{\bar
M}\sqrt{\rho^{(2)}} g^{(1)},\bar M)\big>+\cr&
\big<\sqrt{\rho^{(1)}}g^{(2)},{{\bar M}}^{-1/2}J(\sqrt{\bar M}
\sqrt{\rho^{(1)}}g^{(2)},\bar M)\big>\Big]+\cr
&
{\e}^{-1}\big< \chi{\s} ^{-1} \big[\sqrt{\rho^{(1)}}(K^* h^{(1)}
+ K^*_T h^{(2)})+ {\rho^{(2)}\over\sqrt{\rho^{(1)}}}K^*_G
h^{(1)}\big],g^{(1)}\big>
+ \cr&
{\e}^{-1}\big< \chi{\s} ^{-1} \big[\sqrt{\rho^{(2)}}(K^* h^{(2)}
+ K^*_T h^{(1)})+ {\rho^{(1)}\over\sqrt{\rho^{(2)}}}K^*_G
h^{(2)}\big],g^{(2)}\big>.
}
\Eqa(A.16.1)$$
Here $\big< f,g\big>$ denotes the $L_2(\O\times \Bbb R^3)$ scalar
product. The operator $L$
is symmetric with respect to $\big< \cdot,g\cdot\big>$. The terms in the
first square brackets are non positive by the non positivity of the
operator
$L$. It is easy to see, by using the
symmetry properties of the Boltzmann kernel [CC], that also the
contribution coming
from the terms
in the second square bracket are non positive. Therefore we have
$$
{1\over 2}{d\over dt }||g||_{L_2}^2\le C \ {\e}^{-1} \|h^{(2)})\|_r
\cdot
||g||_{L_2} +\beta
||g||_2{L_2} ^2,
\Eqa(A.18)$$
with $r\ge 3$.
The final estimate is
$$||g||_{L_2}\le C \e^{m-1}|| \nu^{-1} D||_r$$
The same kind of arguments provides the bound for the derivatives of $g$
with respect to 
$x$. Because of the
force terms the argument differs from the one given in [Ca80] in the
fact that we have
to control at the
same time the derivatives with respect to  velocity and space. We
sketch the proof
for the derivatives of
$g$. Differentiating the first equation in \equ(A.9) we get two coupled
equations for
$\nabla_v g$ and
$\nabla g$
$$\eqalign{
\partial_t \nabla g^{(i)} &+(v\cdot \nabla)\n g^{(i)} +\nabla
F^{(i)}\cdot \nabla_v g^{(i)}
+F^{(i)}\cdot
\nabla_v(\nabla g^{(i)})=\nabla
N^{(i)}(g,h)\cr
\partial_t \nabla_v g^{(i)} &+v\cdot \nabla (\nabla_v g^{(i)}) +\nabla
g^{(i)} +F^{(i)}\cdot
\nabla_v(\nabla_v g^{(i)})=\nabla_v
N^{(i)}(g,h) }$$
where $N(g,h)$ is the r.h.s. of \equ(A.9) for $g$.
Proceeding as before in getting \equ(A.16.1) we obtain
$${d\over dt} (||\nabla g||_{L_2}+ ||\nabla_v g||_{L_2})\le c\a_F
(||\nabla
g||_{L_2}+ ||\nabla_v
g||_{L_2}) + ||\nabla N||_{L_2}+
||\nabla_v N||_{L_2}$$

The derivatives of $N$ with respect to velocity  can be estimated by the
methods in
[ELM94], where it is
proven the identity

$${\pt \over \pt v}Q(f,g)=Q (f,{\pt g\over \pt
v}) +Q (g,{\pt f \over \pt v})$$
where ${\pt \over \pt v}$ stands for the partial
derivative with respect to any of
the components of $v$.
The final result is
\goodbreak

{\bf Lemma A.2}

{\it There is an $\e_0>0$ finite for each finite $\a_F$ such that any
solution to the linear
problem
\equ(A.9), with
$D$ and $F_i$ given, satisfies for
$j>3$, $s\le 3$ and any $0<\e\le \e_0$
$$|| g^{(i)}||_{j,s}+|| \partial_v g^{(i)}||_{j,s-1}\le
C(1+\a_F)\e^{m-s}\Big[||\tilde\nu^{-1}
D^{(i)}||_{j+2,s} +||\tilde\nu^{-1}
\partial_v D^{(i)}||_{j+2,s}\Big]
\Eqa(A.19)$$
$$|| h^{(i)}||_{j,s}+|| \partial_v h^{(i)}||_{j,s-1}\le
C(1+\a_F)\e^{m-s+1}\Big[||\tilde\nu^{-1}
D^{(i)}||_{j,s} +||\tilde\nu^{-1}
\partial_v D^{(i)}||_{j,s}\Big]
\Eqa(A.20)$$
}
\vskip.4cm
\noindent{\it Non-linear problem and fixed point argument.}
\vskip.4cm
The nonlinear equations \equ(A.4) are solved by a fixed point method.
This method
works if
there is some small parameter in front of the non linear terms. In
\equ(A.4) there
are two
kinds of non linear terms: the usual Boltzmann non linear term, which is
multiplied by a power of $\e$ and the Vlasov term involving the forces
which gives
rise to
linear and quadratic terms in the remainders: in fact the forces are
given by expressions of
the type
$$F^{(1)}={\bf K}\bstar f^{(2)}={\bf K}\bstar
[\sum_{n=0}^K\e^n f^{(2)}_n]\Big) +\e^m{\bf K}\bstar
R^{(2)}
\Eqa(A.21.0)$$
Hence the non linear term due to the force is small and we can apply the
recursive
argument.

The
Boltzmann terms are dealt with as in Caflisch [Ca80]. The control of
the force
term requires the boundedness of the gradient of the Kac potential. The
result is
\vskip.2cm

\goodbreak

{\bf Theorem A.3}

{\it There is an $\e_0>0$ such that the remainders satisfy for
$j>3$, $z=s-(d-1)$ with $s< m$ and any $0<\e\le \e_0$
$$|| g^{(i)}||_{j,z}\le c \e^{m-s+1},\quad || h^{(i)}||_{j,z}\le c
\e^{m-s+1}, \quad
i=1,2
\Eqa(A.21)$$
}
\vskip.5cm
{\it Proof}. \
Let $R^{(i)}_k$ be the solution of \equ(A.1.2) with
force
$$\eqalign{F^{(i)}_k &=\Big({\bf K}\bstar
[\sum_{n=0}^K\e^n f^{(j)}_n)]\Big) +\e^m\Big({\bf K}\bstar
R^{(j)}_{k-1}\Big) \cr &
:= F_\e^{(i)} +\e^m\tilde F^{(i)}_k, \quad j=i+1\, \hbox{\rm mod }2,
}
\Eqa(A.21.1)$$
and the collision integrals computed with $R^{(i)}$ replaced by
$R^{(i)}_{k-1}$; moreover, we
put $R^{(i)}_0 =0, i=1,2$. Lemma A.2 and an inductive argument assure
that the sequences
$R^{(i)}_k$ are uniformly bounded for $\e$ sufficiently small. In
fact, setting
$\bar\a=\a_{F_\e^{(i)}}$, we have $\a_k=\a_{F^{(i)}_k}\le\bar\a +
C\e^m\|R_{k-1}\|_1$. By a
standard argument $\|R_{k-1}\|_1$ is bounded by a constant
$\lambda(\a_{k-1})$ with
$\l(\,\cdot\,)$ some monotone function. Hence,
$\a=\sup\a_k$ satisfies the inequality $\a\le\bar\a +\e^m\lambda(\a)$.
By setting
$\e^m\lambda(2\bar\a)\le \bar\a$, we conclude $\a<2\bar\a$ and hence the
uniform boundedness
of the sequences $R^{(i)}_k$ for $\e$ sufficiently small.

The differences $\d
R^{(i)}_k:=R^{(i)}_k-R^{(i)}_{k-1}$ can be decomposed again in high
and low
velocity parts $\d g_k, \d h_k$, with $\d g_k=(\d g^{(1)}_k,\d
g^{(2)}_k)$ and
$\d h_k=(\d h^{(1)}_k,\d h^{(2)}_k)$ solve the equations

$$D_t \d g_k +(\e^m\tilde F_{k-1}+F_\e)\cdot\nabla_v\d g_k
= N(\delta g_k ,\d h_k)$$
$$D_t \d h_k +(\e^m\tilde F_{k-1}+F_\e)\cdot\nabla_v\d h_k
= N'(\delta g_k ,\d h_k)$$
where $N=(N^{(1)},N^{(2)})$, $N'=(N'^{(1)},N'^{(2)})$ and
$ N^{(1)}$, $N'^{(1)}$ are the r.h.s. in \equ(A.9) with
$$\eqalign{&D^{(i)}=-\e\d \tilde F_k\cdot\nabla_v ( g_{k-1}+
h_{k-1})+\cr&[J(R^{(i)}_{k-1},R^{(i)}_{k-1})-J(R^{(i)}_{k-2},R^{(i)}_{k-2})]+
[J(R^{(i)}_{k-1},R^{(j)}_{k-1})-J(R^{(i)}_{k-2},R^{(j)}_{k-2})].}$$
with $\d \tilde F_k^{(i)}:=(\tilde F^{(i)}_k-\tilde F^{(i)}_{k-1})={\bf
K}\bstar[R^{(j)}_{k-1}-R^{(j)}_{k-2}]$. By Lemma A.2 the solutions
satisfy \equ(A.19) and
\equ(A.20), so that
$$|| \delta g_k||_{j,s}\le C \e^{m-s+1}||\tilde\nu^{-1}
[\d \tilde F_k)\cdot\nabla_v ( g_{k-1}+ h_{k-1})]||_{j+2,s}
\Eqa(A.19.1)$$
$$|| \d h_k||_{j,s}\le \e^{m-s+2}||\tilde\nu^{-1}
[\d \tilde F_k\cdot\nabla_v ( g_{k-1}+ h_{k-1})]||_{j,s}
\Eqa(A.20.1)$$

We have
$$\eqalign{
||\d \tilde F_k\cdot \nabla_v& ( g_{k-1}+ h_{k-1})||^2_{j0}\le
C\sup_{x,v}(1+|v|^2)^j|\nabla_v (\d g_k+\d
h_k)|
\sup_v\int dx|\d \tilde F_k|^2 \cr &
\le C\int dx\Big|\int dv\int dy K(x-y) (R_{k-1}-R_{k-2})\Big|^2\cr &\le
C\int dx\Big|\sup_v(1+|v|^2)^j\int dy K(x-y) (\d g_{k-1}+\d
h_{k-1})\Big|^2\cr
&\le C
\Big|(\int dy |K(x-y)|^2)^{1/2}\sup_v(1+|v|^2)^j \big(\int dy|\d
g_{k-1}+\d
h_{k-1}|^2\big)^{1/2}\Big|^2\cr &
\le C [||\d g_{k-1}||^2_{j0} +||\d h_{k-1}||^2_{j0}]
}$$
To get the second inequality we have used that the norms $||\nabla_v\d
g_k||_{j,s}$ and
$||\nabla_v\d
h_k||_{j,s}$ are finite by Lemma A.2, so that the supremum over $x$ in
the first row
exists finite. The last inequality is a consequence of the fact that
the
Kac potential is bounded and
that the space integration is on a torus. Finally, by
\equ(A.19.1) and
\equ(A.20.1) we have
$$|| \delta g_k||_{j,0}\le c\e^{m-s+1}||\d g_{k-1}||_{j,s} +||\d
h_{k-1}||_{j,0}]
$$
$$|| \d h_k||_{j,0}\le c\e^{m-s+1}||\d g_{k-1}||_{j,s} +||\d
h_{k-1}||_{j,0}]$$

We remark that it is possible to prove that the norm $||\cdot||_{j,z}$
for the
remainders $g^{(i)}, h^{(i)}$
are bounded with $z=s-(d-1)$ and $s< m$.
\qed
\vskip.4cm
\numfor=1

{\bf Appendix B}
\vskip.3cm
In this Appendix we show how to bound the remainders which are
solutions of
\equ(6.20). The
method we use is
different from the one in Appendix A. In fact in this case the lowest
order is a global
Maxwellian and we do not need to introduce the decomposition into low
and
high
velocity.
Also in this
case we need a Theorem on the regularity of the terms of the expansion
analogous to
Theorem A.1

{\bf Theorem B.1}

{\it Given $\e>0$, assume that there exists a sufficiently smooth
solution of the
incompressible
Navier-Stokes equations
\equ(10) in $(0, t_0]$. Then there is a constant $c>0$ and $s$ depending
on the
smoothness of the
solution of INS such that the terms in the expansion
$f_i,\phi_i,
\ i=2,...,K$ solutions of the equations \equ(6.2.2) and \equ(6.2.3)
satisfy
$$||f_i||_{j,s}\le c,\quad ||\phi_i||_{j,2}\le c
\Eqb(B.0)$$
$$||\partial_vf_i||_{j,s}\le c,\quad ||\partial_v\phi_i||_{j,2}\le
c\Eqb(B.1.1)$$
for any $j$.}
\vskip.2cm
We write \equ(6.20) for the variables $R^{(i)}, 1=1,2$ defined in
\equ(RR)
$$\eqalign{
\pt_t R^{(1)}& + \e^{-1}\big[ v\cdot R^{(1)}+F^{(1)}\cdot\n_v
R^{(1)}\big]=\e^{-2}
\Big[{\cal
L}R^{(1)}+ \Gamma R^{(1)}+\Theta R^{(2)} \Big]\cr &
+\e^{-1}\Big[
{\cal L}_1 R^{(1)}
+ \Gamma_2 R^{(1)} +\Theta_1
R^{(2)} \Big]+ \Big[{\cal L}'_1 R^{(1)}
+ \Gamma'_2 R^{(1)} +\Theta'_1
R^{(2)} \Big] \cr & +
\e^{m-2}\big[J(R^{(1)},R^{(1)})+ J(R^{(1)},R^{(2)})+A^{(1)}\big],\cr
\pt_t R^{(2)}& +\e^{-1}\big[ v\cdot R^{(2)}+ F^2\cdot\n_v R^{(2)}]=
\e^{-2} \Big[ {\cal L}R^{(2)}+ \Gamma R^{(2)}+\Theta
R^{(1)}\Big] \cr &
+\e^{-1}\Big[
{\cal
L}_2 R^{(2)}
+ \Gamma_1 R^{(2)} +\Theta_2 R^{(1)}\Big]+\Big[ {\cal
L}'_2 R^{(2)}
+ \Gamma_1' R^{(2)} +\Theta_2' R^{(1)}\Big]\cr &+
\e^{m-2}\big[J(R^{(2)},R^{(2)})+ J(R^{(2)},R^{(1)})+A^{(2)}\big],
}\Eqb(B.1)$$
$\Theta, {\cal L}, \Gamma$ are defined as $\bar\Theta, \bar{\cal L},
\bar\Gamma$ in
the list after
\equ(A.1.2)
after substituting the global Maxwellian
$M_0(\bar\rho,\bar T)$ to
$\bar M$. Finally,
$${\cal L}_i g =
J(f^{(i)}_1,g)+J(g,f^{(i)}_1)], \quad
\Theta_i g= J(
\f^{(i)}_1, g),\quad \G _i g = J(g,f^{(i)}_1),$$
$${\cal L}'_i g = \sum_{h=2}^K
\e^{h-2}[J(f^{(i)}_h,g)+J(g,f^{(i)}_h)],$$
$$\Theta'_i g= J(\sum_{n=2}^K\e^{n-2}
\f^{(i)}_n, g),\quad \G '_i g = \sum_{h=2}^K \e^{h-2}J(g,f^{(i)}_h),$$
The first step is to consider the linear problem associated to
\equ(B.1), namely to
study \equ(B.1) with
the last terms
$D^{(i)}:=J(R^{(i)},R^{(i)})+
J(R^{(i)},R^{(j)})+A^{(i)}, i=1,2, i\ne j$ given and $F^{(1)}, F^{(2)}$
fixed, independent of
$R^{(1)}, R^{(2)}$. Moreover remembering that the forces vanish to the
lowest order in $\e$,
we assume that the $L_\infty$ norms of $F^{(i)}$ and their gradients are
bounded by some
constant $\e \a_F$. The role of the constant $\a_F$ is similar to the
one discussed in the
previous appendix and we do not repeat the iterative argument in this
case. We will
just provide an estimate for the
$L^2$ norm in $(x,v)$
$||R||_2:=||R^{(1)}||_2 +||R^{(2)}||_2$ for the solution
$R=(R^{(1)},R^{(2)})$ of this
problem, the rest of the argument being standard (see for example
[ELM98]).
We put
$R^{(i)}=\sqrt{M_0} \Psi^{(i)}$ so that
$$\eqalign{
\pt_t \Psi^{(1)}& + \e^{-1}\big[ v\cdot \n \Psi^{(1)}+F^{(1)}\cdot\n_v
\Psi^{(1)}-{1\over
2}\Psi^{(1)}F^{(1)}\cdot v
\big]=\cr &\e^{-2} \Big[ L\Psi^{(1)}+ G \Psi^{(1)}+T \Psi^{(2)} \Big]
+\e^{-1}\Big[
L_1 R^{(1)}
+ G_2 \Psi^{(1)} +T_1
\Psi^{(2)} \Big]+ \cr &\Big[ L'_1 \Psi^{(1)}
+ G'_2 \Psi^{(1)} +T'_1
\Psi^{(2)} \Big] +
\e^{m-2}{D^{(1)}\over \sqrt{M_0}}
}
\Eqb(B.9)$$
where the relation between the old operators ${\cal L}, \Theta, \Gamma,
{\cal L}_i,
\Theta_i, \Gamma_i, {\cal
L}'i, \Theta'_i, \Gamma'_i$ and the new ones $L,T, G, L_i,T_i,G_i,
L'_i,T'_i,G'_i$ is
of the form
$${\cal L}f={1\over \sqrt{M_0}}{ L}\sqrt{M_0}f.$$
It is easy to see that, setting
$\|\Psi\|_{L_2}^2=\|\Psi^{(1)}\|_{L_2}^2+\|\Psi^{(2)}\|_{L_2}^2$,
$$\eqalign{
{1\over 2}{d\over dt} ||\Psi||_{L_2}^2&=\e^{-2} \Big[\big<\Psi^{(1)},L
\Psi^{(1)}\big>+\big<\Psi^{(2)}, L\Psi^{(2)}\big>-{1\over
2}\e\sum_{i=1}^2 F^{(i)}\cdot\big<\Psi^{(i)}, v\Psi^{(i)}\big>\Big]\cr &
+ \e^{-2}\Big[\big< \Psi^{(1)}, M_0^{-1/2}J(M_0,\sqrt{M_0}\
\Psi^{(2)})\big>+\big<
\Psi^{(1)}, M_0^{-1/2}J(\sqrt{M_0}\
\Psi^{(1)},M_0)\big> +\cr &
\big<\Psi^{(2)}, M_0^{-1/2}J(M_0,\sqrt{M_0}\
\Psi^{(1)})\big>+\big<\Psi^{(2)},
M_0^{-1/2}J(\sqrt{M_0}\
\Psi^{(2)},M_0)\big>\Big]\cr &
+\e^{-1}\Big[\big<\Psi^{(1)},L_1 \Psi^{(1)}\big>+\big<\Psi^{(2)},
L_2\Psi^{(2)}\big>
+\big<\Psi^{(1)},(G_2 \Psi^{(1)} +T_1
\Psi^{(2)})\big>+\cr &
\big<\Psi^{(2)},(G_1 \Psi^{(2)} +T_2
\Psi^{(1)})\big>\Big]
+ \Big[\big<\Psi^{(1)},L'_1 \Psi^{(1)}\big>+\big<\Psi^{(2)},
L'_2\Psi^{(2)}\big> \cr &+
\big<\Psi^{(1)},(G'_2 \Psi^{(1)} +T'_1
\Psi^{(2)})\big>+
\big<\Psi^{(2)},(G'_1 \Psi^{(2)} +T'_2
\Psi^{(1)})\big>\Big]\cr &
+\e^{m-2}\sum_{i=1}^2\big<\Psi^{(i)},{D^{(i)}\over \sqrt{M_0}}\big>
}
\Eqb(B.10)$$
First of all, we observe that the terms in the second square bracket are
non positive
[CC].
To estimate the other terms we will use the strict negativity of the
operator $L$
(see \equ(3.14)) on the space orthogonal to the collision invariants
and the following
estimate on the
operator $J(f,h)$ (see for example [GPS]): for any Maxwellian
$M$ and for any $y\in[-1,1]$
$$\int_{\Bbb R^3}\ dv
{|J (\sqrt M f,\sqrt M h)|^2\over \nu M}
\le \int_{\Bbb R^3}\ dv \nu |f|^2\ \int_{\Bbb R^3}\ dv \nu|h|^2
\Eqb(4.24.1)$$
This inequality and the bounds on the $f_n$'s imply the following
bounds:
$$|\big<\sum_{i=1}^2\Psi^{(i)} { L}_i\Psi^{(i)}\big>|\le C\pa
\sum_{i=1}^2\sqrt{\nu }\bar
\Psi^{(i)}\pa_{L_2} \ \pa
\Psi^{(i)}\pa_{L_2} \pa \ M_0^{-1/2}f_1^{(i)}\pa_{j,s},\Eqb(BL1)$$
$$|\big<\sum_{i=1}^2\Psi^{(i)} { L}'_i\Psi^{(i)}\big>|\le C\pa
\sqrt{\nu}
\bar\Psi\pa_{L_2} \ \pa
\Psi\pa_{L_2} \pa M^{-1/2}_0\sum_{n=2}^7
f_n\pa_{j,s}\Eqb(BL2)$$
where $\bar g$ denotes the projection of a function $g$ on the
orthogonal to the invariant
space of $L$, while $\hat g$ is the projection on the invariant space.
Note that
the presence of the product
$\pa\sqrt{\nu}\bar\Psi\pa_2\/\/
\pa\Psi\pa_2$ depends on the fact that $ L_i$ and $ L'_i$ are both
orthogonal to the collision invariants.

Similar estimates hold for the terms involving the other operators. By
using the
bounds on the $f_n$'s and
$\phi_n$'s and after some algebra, the terms in the forth, fifth and
sixth rows are
bounded by
$$C\pa \sqrt{\nu}
\bar\Psi\pa_{L_2} \ \pa
\Psi\pa_{L_2} $$
To bound the last term in the first square bracket of \equ(B.10), we
note
that
$$\e^{-1}|\sum_{i=1}^2 F^{(i)}\cdot\big<\Psi^{(i)}, v\Psi^{(i)}\big>|\le
\a_F[\|\sqrt{\nu}\bar\Psi\|^2_{L_2}+ C\|\hat\Psi\|_{L_2}^2,$$
and we assume $\e$ so small that $\e^2\a_F<1/2$.

We integrate \equ(B.10) in time between $0$ and $ t_0$. With the
notation
$\Psi_t(\,\cdot\,)=\Psi(\,\cdot\,,t)$, we get

$$\eqalign{&{1\over 2} \pa \Psi_{t_0} \pa^2_{L_2} \le
C\int_0^{t_0}d\/t\Bigg\{-\e^{-2}\Big[{1\over 2}\pa
\sqrt\nu\bar \Psi_t\pa^2_{L_2} +C_F\pa
\sqrt\nu \Psi_t\pa^2_{L_2}\Big]\cr & + C(\e^{-1}+1)\pa \sqrt\nu \bar
\Psi_t\pa_{L_2} \ \pa
\Psi_t\pa_{L_2}
+\e^{m-2}\pa
D(\,\cdot\,,t)\pa^2_{L_2}\Bigg\} }\Eqb(B.3)
$$
The first
term in the second line is due to the bounds \equ(BL1) and \equ(BL2).
Moreover

\noindent$\pa M^{-1/2}_0 f_1\pa_{j,s}$ and $ \pa M^{-1/2}_0
f_1\pa_{j,s}$ are
bounded by
the regularity
of the solutions of the macroscopic equations for $0<t< T_0$ and $\pa
M^{-1/2}_0\sum_{n=2}^K f_n\pa_{j,s}\le C$ by Theorem B.1.

Using the inequality
$$ -{1\over\e^{2}} x^2 +(c_1\e^{-1}+c_2)x y \le (c_1+c_2\e)^2y^2/4$$
valid for any positive $\e$, $x$, $y$.
with $x=||\sqrt{\nu}\bar\Psi||_2$, $y=||\Psi||$ and suitable
constants $c_1$ and $c_2$, we get (since $\Psi(\cdot,0)=0$)
$$\eqalign{&\pa \Psi(\,\cdot\,,t_0)\pa^2_{L_2} \le\int_0^{t_0}d\/t
C_F\Bigg[\pa\Psi(\,\cdot\,,t)\pa^2_{L_2} +\pa M_0^{-1/2} D(\,\cdot\,,t)
\pa^2_{L_2}
\Bigg] }$$
In conclusion, by the use of the Gronwall lemma, for $\e$ sufficiently
small, we get:
$$\sup_{0\le t\le t_0} \pa \Psi(\,\cdot\,,t)\pa_{L_2}\le C({t_0}, \a_F)
\sup_{t\in(0,t_0]}
\pa M_0^{-1/2} D(\,\cdot\,,t) \pa_{L_2}\ \Eqb(4.25.1)$$
\vskip.3cm

The bounds for the Sobolev norm of higher order in $x,v$ are obtained by
studying the
equations for the
derivatives as explained in Appendix A. Finally, writing the equations
for the
remainders in the
integral form and using the property of the linearized Boltzmann
operators of
improving the regularity in
$v$, we get the analogous of Lemma A.2
\goodbreak
{\bf Lemma B.2}

\noindent {\it There is an $\e_0>0$ such that any solution to the linear
problem
\equ(B.9), with
$D$ and $F_i$ given, after choosing
$K=2m$ satisfies for
$j>3$, $s<s_0$ and any $0<\e\le \e_0$
$$|| g^{(i)}||_{j,s}+|| \partial_v g^{(i)}||_{j,s-1}\le
\e^{m-2}C_F\Big[||
D^{(i)}||_{j+2,s}+||\tilde\nu^{-1}
\partial_v D^{(i)}||_{j+2,s}\Big]
\Eqb(B.19)$$
$$|| h^{(i)}||_{j,s}+|| \partial_v h^{(i)}||_{j,s-1}\le
\e^{m-2}C_F\Big[||
D^{(i)}||_{j,s}+||\tilde\nu^{-1}
\partial_v D^{(i)}||_{j,s}\Big]
\Eqb(B.20)$$
}

The dependence on the force in the bounds \equ(B.19), \equ(B.20) does
not affect the
argument given in
Appendix A to solve the non-linear problem, because in the bounds for
$R_k$ the
constant $C_{F_k}$ will
depend on the norm of $R_{k-1}$.

\vskip .4cm
\numfor=1

{\bf Appendix C}
\vskip .5cm
To show formally the convergence of the microscopic one particle
distribution
functions to the solution of the VBE in the Grad-Boltzmann limit, let
us
consider the
hierarchy for the rescaled correlation functions
$r_{j_r, j_b}$ of $j_r$ particles of species $r$ and $j_b$ of species
$b$,
defined as
$$\eqalign{&
r_{j_r, j_b}(z_1^r,\cdots, z_{j_r}^r;z_1^b,\cdots, z_{j_b}^b;\t)
=
\d^{-(j_r+j_b)}{N_r!\over (N_r-j_r)!}{N_b!\over
(N_b-j_b)!}\int_{(\L\times
\IR^3)^{(N-j_r-j_b)}}\cr & dz^r_{j_r+1}\cdots dz^r_{N_r}
dz^b_{j_b+1}\cdots dz^b_{N_b}
\mu_N(\d^{-1}q^r_1, v^r_1\cdots,
\d^{-1}q^r_{j_r}, v^r_{j_r};
\d^{-1}q^b_1,v^b_1\cdots,
\d^{-1}q^b_{j_b};\d^{-1}\t))}
\Eqc(0.4)$$
where $z^\a=(q^\a,v^\a)$ is the phase space point of a particle of
species $\a$ and
$\mu_N$ is the probability distribution solution of the Liouville
equation

$$\partial_{\tau_{m}}
\mu_N+\sum_{i=1}^N
v_i\cdot\nabla_{\x_i}\mu_N-A_\ell\sum_{\a\sneq\b}\sum_{i=1}^{N_\a}\sum_{j=1}^{N_\b}
\nabla_{\x^\a_i}U_\ell(|\x^\a_i-\x^\b_j|)\cdot
\nabla_{v_i^\a}\mu_N=0,$$
which is valid in $\G_N$, i.e. where the hard spheres do not overlap.
On the
boundary of $\G_N$ we assume the boundary conditions
$$
\mu_N(\x_1,v_1,\cdots, \x_N, v_N;{\tau_{m}})=
\mu_N(\x_1,v_1,\cdots,\x_i,v'_i,\cdots,\x_j,v'_j,\cdots, \x_N,
v_N;{\tau_{m}})$$
if
$$|\x_i-\x_j|=1,\qquad i\ne j,$$
where $v_i'=v_i-\omega[\omega\cdot(v_i-v_j)], v'_j=
v_j+\omega[\omega\cdot(v_i-v_j)]$ with $\omega$ the unit vector directed
as $\x_i-\x_j$. The above conditions merely state the conservation of
the
probability
during an
elastic collision. As pointed out before, contacts of more than two
particles have
null
Lebesgue measure, so they do not affect above definition.
\vskip.1cm
The rescaled correlation functions satisfy a hierarchy of equations of
the form
$$\eqalign{\pt_\t
r_{j_r,j_b}+&\sum_{i=1}^{j_r}\Big[v^r_i\cdot\n_{q^r_i}r_{j_r,j_b}+
\d^3\sum_{j=1}^{j_b}\n_{q^r_i}V_\gamma(|q_i^r-q_j^b|)\cdot\n_{v^r_i}r_{j_r,j_b}\Big]+\cr&
\sum_{i=1}^{j_b}\Big[v^b_i\cdot\n_{q^b_i}r_{j_r,j_b}+
\d^3\sum_{j=1}^{j_r}\n_{q^b_i}V_\gamma(|q^b_i-q_j^r|)\cdot\n_{v^b_i}r_{j_r,j_b}\Big]=\cr&
\sum_{i=1}^{j_r}\Big[{\cal
B}^{r,r}_{\d,i}r_{j_r+1,j_b}+
{\cal B}^{b,r}_{\d,i}r_{j_r,j_b+1}+{\cal
V}^{b,r}_{i}r_{j_r,j_b+1}\Big] +\cr&\sum_{i=1}^{j_b}\Big[{\cal
B}^{b,b}_{\d,i}r_{j_r,j_b+1}+
{\cal B}^{r,b}_{\d,i}r_{j_r+1,j_b}+
{\cal V}^{r,b}_{i}r_{j_r+1,j_b}\Big],}\Eqc(Hier)$$
where, with the notation $\un z_k=(z_1,\dots,z_k)$, we have
$$\eqalign{&({\cal B}^{r,r}_{\d,i}r_{j_r+1,j_b})(\un z^r_{j_r};\un
z^b_{j_b})=\cr&\int_{\IR^3}
dv^r_{j_r+1}\int_{S^2_+} d\o(v^r_{j_r+1}-v^r_i)\cdot \o
\Big [r_{j_r+1,j_b}(\un
z_{j_r+1}^r)';\un z^b_{j_b})-r_{j_r+1,j_b}(\overline{\un
z^r_{j_r+1}};\un
z^b_{j_b})\Big],}$$
with
$$S^2_+=\{\o\in \IR ^3\,\,|\,\, |\o|=1,\,\, \o\cdot
(v^r_{j_r+1}-v^r_i)>0\},$$
$$(\un z_{j_r+1}^r)'= (z^r_1,\dots (z^r_i)', \dots,
z^r_{j_r},(z^r_{j_r+1})'),$$
$$\overline{\un z^r_{j_r+1}}= (z^r_1,\dots \bar z^r_i, \dots,z^r_{j_r},
\bar
z^r_{j_r+1})$$
and, for any
$z$, $z_*$, the phase points
$z'$, $z'_*$, $\bar z$ and $\bar z_*$are defined by
$$q'=q, \quad q'_*=q+\d \o,\quad v'=v-\o(\o\cdot(v-v_*),\quad
v'_*=v_*+\o(\o\cdot(v-v_*),$$
$$\bar q=q, \quad \bar q_*=q-\d \o,\quad \bar v=v,\quad
\bar v_*=v_*.$$
Moreover,
$$\eqalign{&({\cal B}^{b,r}_{\d,i}r_{j_r,j_b+1})(\un z^r_{j_r};\un
z^b_{j_b})=\cr&\int_{\IR^3}
dv^b_{j_r+1}\int_{S^2_+} d\o(v^b_{j_b+1}-v^r_i)\cdot \o
\Big [r_{j_r,j_b+1}((\un z_{j_r}^r)';(\un
z^b_{j_b+1})')-r_{j_r,j_b+1}(\overline{\un
z^r_{j_r}};\overline{\un z^b_{j_b+1}})\Big],}$$
where
$$(\un z_{j_r}^r)'=(z^r_1,\dots (z^r_i)', \dots, z^r_{j_r}),\quad (\un
z^b_{j_b+1})'=
(z^b_1,\dots,z^b_{j_b}, (z^b_{j_b+1})'),$$
$$\overline{\un z_{j_r}^r}=(z^r_1,\dots \bar z^r_i, \dots,
z^r_{j_r}),\quad
\overline{\un
z^b_{j_b+1}}= (z^b_1,\dots,z^b_{j_b}, \bar z^b_{j_b+1}).$$
The collision terms ${\cal B}^{b,b}_{\d,i}$ and ${\cal B}^{r,b}_{\d,i}$
are defined in
a similar
way. Furthermore
$$\eqalign{&({\cal V}^{b,r}_{i}r_{j_r,j_b+1})(\un z^r_{j_r};\un
z^b_{j_b})=
\cr&-\int_{\IR^3}
dv^b_{j_r+1}\int_{\L}
dq^b_{j_b+1}\gamma^3\n_{q^r_i}U_\g(|q_i^r-q^b_{j_b+1}|)\n_{v^r_i}
r_{j_r,j_b+1}(\un z_{j_r}^r;\un z^b_{j_b+1})}
$$
and a similar expression for ${\cal V}^{r,b}_{i}$.
Taking formally the limit $\d\to 0$, the limiting correlations satisfy
the following
Vlasov-Boltzmann hierarchy:
$$\eqalign{&\pt_\t
r_{j_r,j_b}+\sum_{i=1}^{j_r}v^r_i\cdot\n_{q^r_i}r_{j_r,j_b}+
\sum_{i=1}^{j_b}v^b_i\cdot\n_{q^b_i}r_{j_b,j_b}=\cr&
\sum_{i=1}^{j_r}\Big[{\cal B}^{r,r}_{i}r_{j_r+1,j_b}+
{\cal B}^{b,r}_{i}r_{j_r,j_b+1}+{\cal V}^{b,r}_{i}r_{j_r,j_b+1}
\Big]+\cr&
\sum_{i=1}^{j_b}\Big[{\cal B}^{b,b}_{i}r_{j_r,j_b+1}+
{\cal B}^{r,b}_{i}r_{j_r+1,j_b}+
{\cal V}^{r,b}_{i}r_{j_r+1,j_b}\Big],}\Eqc(BHier)$$
where,
$$\eqalign{&({\cal B}^{r,r}_{i}r_{j_r+1,j_b})(\un z^r_{j_r};\un
z^b_{j_b})=\cr&\int_{\IR^3}
dv^r_{j_r+1}\int_{S^2_+} d\o(v^r_{j_r+1}-v^r_i)\cdot \o
\Big [r_{j_r+1,j_b}(\un
z_{j_r+1}^r)';\un z^b_{j_b})-r_{j_r+1,j_b}(\un z^r_{j_r+1};\un
z^b_{j_b})\Big]}$$
and, for any
$z$, $z_*$, the phase points
$z'$, $z'_*$ are defined by
$$q'=q, \quad q'_*=q,\quad v'=v-\o(\o\cdot(v-v_*),\quad
v'_*=v_*+\o(\o\cdot(v-v_*).$$
Similar modifications provide the other terms of the Vlasov-Boltzmann
hierarchy.

It is easy to see that if the initial condition of {\it molecular
chaos}
$$r_{j_r,j_b}(\un z^r_{j_r};\un
z^b_{j_b};0)=\prod_{i=1}^{j_r}f^{r}(z^r_i,0)
\prod_{k=1}^{j_b}f^{b}(z^b_k,0)$$
\noindent is satisfied, then the correlation functions stay factorized
at positive
times
$\t$ and
$f^{r}(q,v,\t)$ and $f^{b}(q,v,\t)$ are the solutions of the coupled
Vlasov-Boltzmann
equations

$$\eqalign{
\partial_\t &f^r(q,v,\t) +v\cdot\nabla_q f^r(q,v,\t)+F^r \cdot \n_v\
f^r(q,v,\t)
= J(f^r, f^r+f^b),\cr
\partial_\t &f^b(q,v,\t) +v\cdot\nabla_q f^b(q,v,\t)+F^b \cdot \n_v\
f^b(q,v,\t)
= J(f^b, f^r+f^b),}\Eqc(0.7)$$
where
$$F^{r}(q,\t)= -\n_q\int_\O d\/q' \gamma^3(\n
U_\g)(|q-q'|)\int_{\IR^3}dv
f^{b}(q',v,\t),\Eqc(0.22)$$
$$F^{b}(q,\t)= -\n_q\int_\O d\/q'\gamma^3(\n U_\g)(|q-q'|)\int_{\IR^3}dv
f^{r}(q',v,\t),\Eqc(0.33)$$ and
$$J(f,g)=\int_{\IR^3} d\/v_*\int_{S^2_+}d\/\omega (v-v_*)\cdot\omega
[f(v')g(v'_*)-f(v)g(v_*)].
\Eqc(0.44)$$
\vskip.3cm
Summarizing, we obtained formally the Vlasov-Boltzmann equations for a
binary mixture,
where the
Boltzmann collision kernel terms are due to the short range interaction
while the
Vlasov
self consistent force is due to the repulsive weak long range
interaction.
If, instead of the hard core interaction the short range force is given
by a finite
range potential,
we would get formally the same equations but with a different cross
section.

We want to stress that an important step is missing in order to make
the above derivation rigorous. The first rigorous result on the
derivation
of the Boltzmann equation has been given by Lanford [Lan] where the
convergence
of the correlation functions is proven in $L_\infty$-norms. On the other
hand, the
derivation of the Vlasov equation is based on the use of the
variation norm and we have not been able to find a norm suited for both
terms. The only related result, as far as we know, has been obtained in
[GM]
and is about a stochastic particle systems
converging
to a Vlasov-Boltzmann equation with a modified Boltzmann kernel
(Povzner).
The proof is based on martingale methods. In the linear case of a Lorentz
gas  with  a Kac potential term it is possible to prove the convergence
to a Boltzmann equation with a linear collision term and a non-linear
self-consistent force term [MR].
\vskip 1cm

{\bf Acknowledgements}: The work on this paper was partially done during
visits
of R. Esposito and R. Marra to DIMACS and Math. Department, Rutgers
University,
and of R.Esposito, J.L.Lebowitz and R. Marra to IHES. We thank those
institutions for their
hospitality.
Research at Rutgers was partially supported by NSF Grant DMR-9813268,
AFOSR Grant F49620-98-1-0207 and
by DIMACS and its supporting agencies, the NSF under contract
STC-91-1999 and the NJ
Commission on Science and Technology. Supported also by the Italian
MURST and INFM.
Part of this work (S.B.) was performed under the auspices of the U. S.
Department of Energy
by University of California Lawrence Livermore National Laboratory under
Contract No. W-7405-Eng-48.

{\leftskip 2cm\rightskip1cm
\vskip1cm\centerline{\bf References}\vskip1cm
\novepunti

\item{[AGA]} F. J. Alexander, A. L. Garcia and B. J. Alder, {\it A
consistent Boltzmann
algorithm}, Phys. Rev. Lett. {\bf 74}, 5212--15 (1996).
\vskip.1cm
\item{[BELMII]} S. Bastea, R. Esposito, J. L. Lebowitz and R.Marra, in
preparation.
\vskip.1cm
\item{[BL]} S. Bastea and J.L. Lebowitz, {\it Spinodal decomposition in
binary
gases.}, Phys.
Rev. Lett. {\bf 78}, 3499 (1997).
\vskip.1cm
\item{[Ca80]} R. Caflisch, {\it The fluid dynamical limit of the
nonlinear Boltzmann equation}, Commun. Pure and Appl. Math.
{\bf 33}, 651--666 (1980).
\vskip.1cm
\item{[Ca87]} R. Caflisch, {\it Asymptotic expansions of solutions for
the Boltzmann
equation.}, Transp. Th. Stat. Phys. {\bf 16}, 701--725 (1987).
\vskip.1cm
\item{[CC]} S. Chapman and T. G. Cowling, {\it The Mathematical
Theory of
Non-uniform
Gases}, Cambridge Univ. Press, Cambridge, England (1970).
\vskip.1cm
\item{[C]} C. Cercignani, {\it On the Boltzmann equation for rigid
spheres},
Transp. Th. Stat. Phys. {\bf 2}, 211 (1972); C. Cercignani, R. Illner,
and M.
Pulvirenti, {\it The Mathematical Theory of Dilute Gases},
Springer-Verlag,
New York (1994).
\vskip.1cm
\item{[CH]} J. W. Cahn and J. I. Hilliard, {\it Free energy of a
nonuniform system I.
Interfacial free energy}, J. Chem. Phys. {\bf 28}, 258 (1958).
\vskip.1cm
\item{[DS]} Luis De Sobrino, {\it On the kinetic theory of a Van der
Waals gas},
J. Can. Phys. {\bf 45}, 363 (1967).
\vskip.1cm
\item{[DE]} M. Di Meo and R. Esposito,
{\it The Navier-Stokes limit of the stationary Boltzmannn
equation for hard potentials},
J. Stat. Phys. {\bf 84}, 859--874 (1996).
\vskip.1cm
\item{[DEL]} A. De Masi, R. Esposito and J. L. Lebowitz, {\it
Incompressible Navier-Stokes and Euler limits of the Boltzmann
equation}, Commun. Pure and Appl. Math. {\bf 42}, 1189--1214 (1989).
\vskip.1cm
\item{[ELM94]} R. Esposito, J. L. Lebowitz and R. Marra,
{\it Hydrodynamic limit of the stationary Boltzmann equation
in a slab}, Commun. Math. Phys. {\bf 160}, 49--80 (1994).
\vskip.1cm
\item{[ELM95]} R. Esposito, J. L. Lebowitz and R. Marra,
{\it The Navier-Stokes limit of stationary solutions
of the nonlinear Boltzmann equation},
J. Stat. Phys. {\bf 78}, 389--412 (1995).
\vskip.1cm
\item{[ELM98]} R. Esposito, J. L. Lebowitz and R. Marra, {\it
Solutions to the Boltzmann equation in the Boussinesq regime},
J. Stat. Phys. {\bf 90}, 1129--1178 (1998).
\vskip.1cm
\item{[ELM99]} R. Esposito, J. L. Lebowitz and R.Marra,
{\it On the derivation of Hydrodynamics from the Boltzmann equation},
Phys. of Fluids {\bf 11}, 2354--2366 (1999).
\vskip.1cm
\item{[FLP]} P. Fratzl, O. Penrose and J. L. Lebowitz,
{\it Modeling of phase separation in alloys with coherent elastic
misfit},
J. Stat. Phys. {\bf 95}, 1429--1503 (1999).
\vskip.1cm
\item{[G]} M. Grmela, {\it Kinetic equation approach to phase
transitions},
J. Stat. Phys. {\bf 3}, 347 (1971).
\vskip.1cm
\item{[GL96]} G. Giacomin and J. L. Lebowitz,
{\it Exact macroscopic description of phase segregation in
model alloys with long range interactions}, Phys.
Rev. Lett. {\bf 76}, 1094 (1996).
\vskip.1cm
\item{[GL97]} G. Giacomin and J. L. Lebowitz
{\it Phase segregation dynamics in particle systems with long range
interaction I.
Macroscopic limits}, J. Stat. Phys. {\bf 87}, 37--61 (1997);
{\it Phase segregation dynamics in particle systems with long range
interaction II.
Interface motion}, SIAM J. Appl. Math. {\bf 58}, 1707--1729
(1998).
\vskip.1cm
\item{[GM]}C. Graham, S. Meleard,
{\it Stochastic particle approximations for generalized Boltzmann models
and
convergence
estimates}, The Annals of probability {\bf 25}, 115--132 (1997).
\vskip.1cm
\item{[{GPS}]}
F. Golse, B. Perthame, C. Sulem,
{\it On a boundary layer problem for the nonlinear Boltzmann
equation}, Arch. Rat. Mech. Anal. {\bf 104}, 81--96 (1988).
\vskip.1cm
\item{[Gra]} H. Grad,{\it Asymptotic theory of the Boltzmann
equation II}, in {\it Rarified Gas Dynamics}, J. A. Laurmann
ed., Vol. I, 26--59 (1963).
\vskip.1cm
\item{[GSS]} J. Gunton, M. San Miguel and P.S. Sahni, in {\it
Phase Transitions and Critical Phenomena}, Vol. 8, C. Domb and
J.L.Lebowitz eds.,
Academic Press, New York (1989).
\vskip.1cm
\item{[L]} J. S. Langer, {\it An introduction to the kinetics of
first-order phase
transitions}, in Solids far from equilibrium, C. Godreche ed., Cambridge
Univ. Press, Cambridge, England (1991).
\vskip.1cm
\item{[Lac]} M. Lachowicz, {\it On the initial layer and the
existence theorem for the nonlinear Boltzmann equation}, Math.
Meth. Appl. Sci. {\bf 9(3)}, 27--70 (1987).
\vskip.1cm
\item{[Lan]} O. E. Lanford III, {\it The evolution of large
classical systems}, in {\it Dynamical Systems, Theory and
Applications}, J. Moser ed., Lect. Notes in Phys. {\bf 35},
1--111, Springer, Berlin (1975).
\vskip.1cm
\item{[LP]}
J. L. Lebowitz and O. Penrose, {\it Rigorous treatment of the Van der
Waals Maxwell
theory of the liquid vapor transition}, J. Math. Phys. 7, 98--113
(1966).
\vskip.1cm
\item{[MR]} R.Marra and V. Ricci, in preparation.
\vskip.1cm
\item{[OP]} Y. Oono and S. Puri, {\it Study of phase-separation dynamics
by use of
cell dynamical system}, Phys. Rev. A {\bf 38}, 434--453 (1988).
\vskip.1cm
\item{[S]} E. D. Siggia, {\it Late stages of spinodal decomposition in
binary
mixtures}, Phys Rev A {\bf 20}, 595--605 (1979).
\vskip.1cm
}
\end